\documentclass[preprint]{emulateapj}
\usepackage[top=1.7in,bottom=-0.2in,left=1.0in,right=0.4in]{geometry}                
\usepackage{graphicx}
\usepackage{amssymb}
\usepackage{epstopdf}
\usepackage{epsfig}
\usepackage{natbib}
\newcommand{\na}{New Astronomy}
\newcommand{\obj}{CXOJ1426+35}
\begin{document}

\submitted{accepted by ApJ}

\shorttitle{A Candidate Dual AGN at $z=1.175$}
\shortauthors{Barrows et al.}

\title{A Candidate Dual AGN at $z=1.175$}
\author{R. Scott Barrows,\altaffilmark{1} Daniel Stern,\altaffilmark{2} Kristin Madsen,\altaffilmark{3} Fiona Harrison,\altaffilmark{3} Roberto J. Assef,\altaffilmark{2} Julia M. Comerford,\altaffilmark{4} Michael C. Cushing,\altaffilmark{2} Christopher D. Fassnacht,\altaffilmark{5} Anthony Gonzalez,\altaffilmark{6} Roger Griffith,\altaffilmark{7} Ryan Hickox,\altaffilmark{8} J. Davy Kirkpatrick,\altaffilmark{7} and David J. Lagattuta\altaffilmark{5}}
\altaffiltext{1}{Arkansas Center for Space and Planetary Sciences, University of Arkansas, Fayetteville, AR 72701\newline \indent \emph{author email: rbarrows@uark.edu}}
\altaffiltext{2}{Jet Propulsion Laboratory, California Institute of Technology, 4800 Oak Grove Drive, Pasadena, CA 91109}
\altaffiltext{3}{California Institute of Technology, 1200 East California Boulevard, Pasadena, CA 91125}
\altaffiltext{4}{Astronomy Department, University of Texas at Austin, Austin, TX 78712}
\altaffiltext{5}{University of California at Davis, One Shields Avenue, Davis, CA 95616}
\altaffiltext{6}{Department of Astronomy, University of Florida, Gainesville, FL 32611}
\altaffiltext{7}{Infrared Processing and Analysis Center, California Institute of Technology, Pasadena CA 91125}
\altaffiltext{8}{Department of Physics, Durham University, Durham DHI 3LE, UK}
\bibliographystyle{apj}


\begin{abstract}
The X-ray source CXOXBJ142607.6+353351 (CXOJ1426+35), which was identified in a 172 ks $Chandra$ image in the Bo\"{o}tes field, shows double-peaked rest-frame optical/UV emission lines, separated by $0.69^{\prime \prime}$ (5.5 kpc) in the spatial dimension and by 690 km s$^{-1}$ in the velocity dimension.  The high excitation lines and emission line ratios indicate both systems are ionized by an AGN continuum, and the double-peaked profile resembles that of candidate dual AGN.  At a redshift of $z=1.175$, this source is the highest redshift candidate dual AGN yet identified.  However, many sources have similar emission line profiles for which other interpretations are favored.  We have analyzed the substantial archival data available in this field, as well as acquired near-infrared (NIR) adaptive optics (AO) imaging and NIR slit spectroscopy.  The X-ray spectrum is hard, implying a column density of several $10^{23}$ cm$^{-2}$.   Though heavily obscured, the source is also one of the brightest in the field, with an absorption-corrected 2-10 keV luminosity of $\sim 10^{45}$ erg s$^{-1}$.  Outflows driven by an accretion disk may produce the double-peaked lines if the central engine accretes near the Eddington limit.  However, we may be seeing the narrow line regions of two AGN following a galactic merger.  While the AO image reveals only a single source, a second AGN would easily be obscured by the significant extinction inferred from the X-ray data.  Understanding the physical processes producing the complex emission line profiles seen in CXOJ1426+35 and related sources is important for interpreting the growing population of dual AGN candidates.
\end{abstract}

\section{Introduction}
\label{intro}

It is now generally believed that galaxies evolve in a hierarchical fashion, with galaxies growing through mergers.  Furthermore, evidence suggests that most, if not all, galaxies contain a supermassive black hole (SMBH) at their center \citep{Richstone1998}.  Therefore, in a galactic merger the force of dynamical friction will act to bring each SMBH toward the center of the new mass distribution, first forming a dual SMBH (separations of several kpc), then a gravitationally bound system (sub-pc separations) and eventually merging to form a single, more massive SMBH.  The amount of time spent in each stage is dependent upon the efficiency of the processes carrying away angular momentum from the system \citep{Begelman1980}.  At large separations ($>10-100$ pc), the primary force acting on the SMBHs is dynamical friction from the surrounding gas and dust.  As the SMBHs form a binary system (i.e. harden), ejection of stars becomes most important.  Finally at close separations ($<10^{-2}$ pc), gravitational radiation will become the most important mechanism of energy loss \citep{Milosavljevic2003}.

Numerical simulations show that in a wet merger (i.e., where the merging galaxies possess a significant fraction of gas and dust), tidal interactions are effective at funneling the gas and dust toward the center of the merging system where the two SMBHs are located, providing fuel for enhanced accretion \citep{Hernquist:1989}.  This process is thought to be a possible mechanism for initiating an active galactic nucleus (AGN) or quasar phase  \citep{Hopkins05,Hopkins2008}, and the existence of a dual AGN (two actively accreting SMBHs within the same galaxy) is a natural prediction of this merger fueling model.  Therefore, identifying pairs of SMBHs in the process of merging is extremely important for probing the link between AGN activity and galaxy evolution.

Unfortunately, observational evidence of such systems has proven very difficult to obtain.
Large samples of binary quasars exist as evidence for quasar clustering and mutual triggering \citep{Hennawi2006,Hennawi2010,Myers2007,Myers2008,Foreman2009}.  However, these are typically at large projected separations ($>10$ kpc) and thus generally only represent the earliest stage of binary SMBH evolution.  Most confirmed cases of dual SMBHs at smaller separations (between 10 kpc and 1 kpc) have been identified in local ultraluminous infrared galaxies (ULIRGs) as pairs of X-ray point sources \citep{Komossa2003,Guainazzi2005,Hudson2006,Bianchi2008,Piconcelli2010,Koss:2011}, with much of the optical emission obscured by the heavy internal extinction common in ULIRGs.  In some cases the obscuration is so heavy that only one nucleus is visible, thus highlighting the difficulty that dusty, merging systems present to identification of dual AGN.
Additionally, the close (7 pc separation) binary SMBH system in the radio galaxy 0402+379 was identified from Very Long Baseline Array (VLBA) observations \citep{Rodriguez:2006}.  While the excellent spatial resolution achievable in the radio regime makes the VLBA capable of detecting close binaries, this technique is time consuming and not suited to identifying a large sample of sources.

In the past few years, several interesting individual examples of dual/binary SMBH candidates have been  identified from single-epoch optical spectra of AGN or quasars showing double-peaked emission lines.  The Sloan Digital Sky Survey (SDSS) spectral archive \citep{Abazajian09} has been a fruitful database for such identifications.  However, these objects do not represent a homogeneous class, and thus the physical interpretation of many of them, based on single-epoch fiber spectra alone, remains ambiguous.  

For example, the sources SDSS J153636.22+044127.0 \citep[J1536+0441,][]{BL09} and SDSS J105041.35+345631.3 \citep[J1050+3456,][]{Shields09} show double-peaked, broad Balmer emission lines (H$\alpha$, H$\beta$ and H$\gamma$), which resemble two broad line regions (BLRs) with significant velocity differences ($> 2000$ km s$^{-1}$).  However, these objects might instead be members of the sub-class of AGN/quasars called `double-peaked emitters ' (also known as `disk-emitters') where the double peaks of the Balmer lines are attributed to emission from a single accretion disk.  Line emission contribution from an accretion disk is suggested by the extended wings in the Balmer emission lines (particularly evident in H$\alpha$, e.g. \citealp{Chornock10}) which is a characteristic disk profile feature.  On the other hand, these sources are extreme examples of `disk emitters,' and it has been shown that such strongly double-peaked Balmer lines in quasars may represent cases in which the presence of a secondary SMBH is effective at enhancing the double-peaked profile produced by the disk \citep{TG09,Barrows:2011}.  One such object, 4C+22.25, identified by \citet{Decarli2010} as a candidate binary SMBH, shows very high (8700 km s$^{-1}$) velocity offsets between the two peaks of the Balmer emission lines; it is therefore unlikely to be a `disk-emitter' as the velocity difference is several thousand km s$^{-1}$ larger than that of any known similar source \citep[see the sample of][]{Strateva03}.  These results demonstrate the diversity of double-peaked broad line profiles and emphasize that in a close binary SMBH there are likely to be strong interactions between the SMBHs and the accretion disks.    

A growing number of objects have been identified with narrow forbidden emission lines that are either offset from the systemic velocity or double-peaked in single-epoch spectra, many of which have been found through systematic searches for candidate dual AGN, both in the DEEP2 Galaxy Redshift Survey \citep{Comerford2009a} and SDSS \citep{Zhou04,Wang2009,Smith2010,Liu2010a}.  Specifically, objects have been identified as dual AGN candidates because they show strongly double-peaked [OIII]5007 lines \citep{Zhou04,Gerke2007,Peng:2011}, or double-peaks in all of the narrow lines, both permitted and forbidden, e.g. SDSS 092712.65+294344.0 \citep[J0927+2943,][]{Komossa08a,Bogdanovic09b,Dotti09}.  These double-peaks indicate two sets of narrow emission line regions (NLRs), and are suggestive of dual AGN with separations up to several kpc.  These systems are excellent sources for study because, at the implied kpc-scale separations, the two putative AGN can be resolved in two-dimensional spectroscopy and imaging.  

However, double-peaked narrow lines have been seen in a number of AGN and quasars where other physical explanations have been favored.  For example, NLRs are thought to have complex gas kinematics which may include radially flowing material capable of producing offset or double-peaked narrow emission lines \citep{Heckman1981}.  In particular, powerful radio galaxies often show complex, spatially extended NLRs with multiple components aligned along the radio axis (see reviews of radio galaxies in \citealp{McCarthy1993} and \citealp{Miley:DeBreuck:2008}).  Models for this alignment include reflection of the AGN emission \citep{Tadhunter:1988}, jet-induced star formation \citep{DeYoung:1981}, inverse Compton scattering of cosmic microwave background photons off the relativistic electrons in the radio jet \citep{Daly:1992}, material out-flowing from an accetion disk or entrained in a radio jet \citep{Holt2003,Holt2008,Komossa2008b}, as well as a combination of shocks and photoionization by the central engine \citep{Maxfield2002}.  There is an extensive literature on the spatially extended, kinematically complex NLRs of powerful radio galaxies \citep{vanOjik:1996,Villar-Martin:2000,Villar-Martin:2007,Villar-Martin:2010,Zirm:2005,Nesvadba:2006}.  Though radio galaxy NLRs often appear vaguely reminiscent of dual AGN candidates, multiple AGN are rarely suggested as possible interpretations of the complex radio galaxy NLRs.  Finally, simple geometrical interpretations have been invoked to explain double-peaked narrow lines in AGN, such as a rotating, disk-shaped NLR or a single AGN illuminating two NLRs \citep{XK09}. 

Developing a clear understanding of the physical process producing these often complex emission line profiles is important to both the search for dual AGN and to further understand quasar/AGN emission line regions in general.  Since it is clear that a number of physical processes other than a pair of AGN can explain double-peaked emission lines (broad and/or narrow), more information is necessary to entirely reject or confirm any one model.  From single-epoch spectra, emission line ratios may be used as diagnostic tools to constrain the nature of the ionizing continuum producing each set of emission lines.  Other observational tools include spatially resolved spectroscopy, synoptic spectroscopic observations, and high resolution imaging. 

In this paper we discuss the X-ray source CXOXBJ142607.6+353351 (from here on CXOJ1426+35) which was originally identified as an X-ray source in the Bo\"{o}tes field \citep{Wang04}.  In follow-up slit-spectroscopy, CXOJ1426+35 showed double rest-frame optical/UV emission lines, separated spatially by $0.69^{\prime \prime}$, and in velocity-space by 690 km s$^{-1}$, as measured from the [OIII]5007 line.  We discuss CXOJ1426+35 as an example of a candidate dual AGN, and place its spectral and morphological properties within the context of other candidate dual/binary SMBHs.  We examine a variety of possible explanations for the double-peaked emission lines.  Interestingly, CXOJ1426+35 is at $z=1.175$, the highest redshift yet for a candidate dual AGN, where previously candidates were identified out to $z\approx 0.8$.  At this redshift, we have access to rest-frame UV emission lines which is unique among the current sample of candidate dual AGN in the literature.  Section \ref{properties} discusses the known properties of CXOJ1426+35 and Section \ref{interpretation} discusses various physical interpretations of the data.  Section \ref{tests} takes CXOJ1426+35 as a pedagogical example to describe a number of methods by which candidate dual/binary SMBHs may be confirmed or rejected.  Section \ref{conclusions} contains the conclusions.  Throughout the paper we assume cosmological parameters of $H_0=70$ km s$^{-1}$ kpc$^{-1}$, $\Omega_{M}=0.30$ and $\Omega_{\Lambda}=0.70$.   

\section{Properties of the Source}
\label{properties}

In an attempt to understand the nature of the complex line emission of CXOJ1426+35, we first summarize in this section the known properties of the source.  The multi-wavelength data available for CXOJ1426+35 include both archival data and newly obtained data from the Keck II telescope in Hawaii.   

\subsection{Archival Data}
\label{archival}

The photometric properties of CXOJ1426+35 from archival observations are summarized in Table \ref{phot}.  The mid-IR through near-ultraviolet (NUV) colors are displayed as a broad-band spectral energy distribution with best-fitting spectral templates in Section \ref{sed_model}.  We begin by describing the archival observations. 

\subsubsection{Chandra X-ray Data}
\label{Chandra}

CXOJ1426+35 was detected with the Advanced CCD Imaging Spectrometer \citep[ACIS-I;][]{Garmire2003} on board the \emph{Chandra X-Ray Observatory} (CXO) in a 172 ks image of the Large Area Lyman Alpha (LALA) Bo\"{o}tes field \citep{Wang04}.  The total image was composed of two exposures taken on April 16-17 and June 9 of 2002.  In \citet{Wang04}, the reported number of background-subtracted soft (0.5 - 2.0 keV) and hard (2.0-7.0 keV) counts, determined by `wavdetect,' are $24.3^{+6.6}_{-5.4}$ and $240.8^{+17.2}_{-11.0}$, respectively, which yields a hardness ratio (defined as $HR\equiv(H-S)/(H+S)$, where $H$ and $S$ are the number of the hard and soft counts, respectively) of 0.82.  Assuming Galactic absorption and a power law spectrum, the hardness ratio corresponds to a photon index of $\Gamma=-1.3$, which is significantly harder than the typical observed $\Gamma=1.4$ of AGN when not corrected for absorption.  Interestingly, out of 168 sources detected in the field by \citet{Wang04}, only thirteen have larger 0.5-10 keV fluxes and only seven are more obscured (i.e., have higher hardness ratios); no brighter sources have higher hardness ratios.  Thus, CXOJ1426+35 is unique in that it is both relatively bright and heavily obscured \citep[i.e., Figure 2 of][]{Wang04}.

\begin{figure*}
\centerline{\epsfig{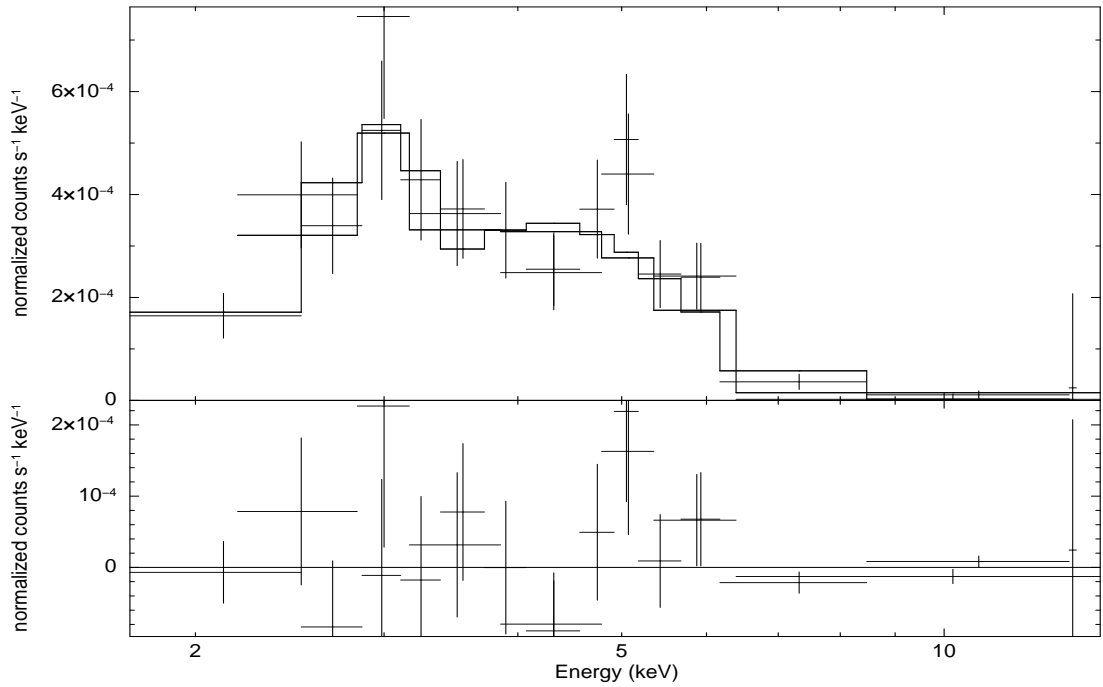}}
\caption{\emph{Chandra}/ACIS spectrum and model residuals of CXOJ1426+35 in the observed frame.  The top panel shows the combined data sets of the extractions from both the 120 ks and 58 ks images (in \citealt{Wang04} the 58 ks image was corrected to 52 ks, hence their quoted total time of 172 ks), and the best fit absorbed power-law model is overlaid.  The bottom panel shows the residuals from the best fit model.}
\label{xspectra}
\end{figure*}

Unfortunately, the off-axis detection of CXOJ1426+35 means that it is unresolved by CXO.  However, the high number of hard counts allowed us to obtain a reasonable spectrum.  To do so, we re-reduced the data using the CIAO software, following the standard procedure for extracting source and background regions, and creating response files for spectral analysis.  The two observations were processed independently.  In each event file, we extracted a source region 30 pixels ($15^{\prime \prime}$) in diameter centered on the centroid position of the source and a background region 50 pixels ($25^{\prime \prime}$) in diameter adjacent to the source region.  The source regions, background regions and response files were used in XSPEC to perform spectral modeling.  For each event file, the spectra were re-binned so that each bin had at least 15 counts.  When fitting the models, the same model was fit to the spectrum from each observation simultaneously to place tighter constraints on the parameters.  All bad channels were ignored, and the first channel of each data set (where there are few counts) was also ignored.  A simple absorbed power law was fit with the redshift frozen at $z=1.175$ (see Section \ref{lris} for the redshift determination).  The spectrum, with the model overlaid and residuals underneath, is shown in Figure \ref{xspectra}.  The residuals around $\sim3$ keV are suggestive of rest-frame 6.4 keV Fe K$\alpha$ emission, though there are too few counts to obtain a good line fit.  The best fit parameters are an intrinsic power-law slope $\Gamma=1.96^{+1.24}_{-0.57}$ and column density $n_{H}=5.90^{+4.94}_{-1.81}\times 10^{23}$ cm$^{-2}$.  This column density implies a highly obscured source.  From the best fit model the observed 2-10 keV flux is $5.44^{+0.65}_{-0.58}\times 10^{-14}$ erg s$^{-1}$ cm$^{-2}$ or $2.81^{+0.33}_{-0.30}\times 10^{-3}$ $\mu$Jy (Table \ref{phot}) which corresponds to an absorption-corrected intrinsic luminosity of $L_{\rm 2-10 keV}=9.7^{+3.0}_{-2.3}\times 10^{44}$ erg s$^{-1}$.  From $L_{\rm 2-10 keV}$, the bolometric correction of \citet{Hopkins2007} yields $L_{\rm BOL}\approx 9\times 10^{46}$ erg s$^{-1}$.

\subsubsection{UV Imaging}
\label{uv}

CXOJ1426+35 was detected by the NUV detector on the \emph{Galaxy Evolution Explorer} \citep[\emph{GALEX};][]{Martin2003} which has an observed bandwidth of 1771-2831 \AA~ and effective wavelength of 2271 \AA.  The $5^{\prime \prime}$ FWHM resolution of the NUV detector on \emph{GALEX} is unable to resolve CXOJ1426+35.  Therefore, for photometry we have used the flux measurement from the 3.8$^{\prime \prime}$ radius aperture (to most closely match the 4$^{\prime \prime}$ radius aperture for the IRAC data; Section \ref{mid-ir}).  We then corrected to a total magnitude using the correction factor from \citet{Morrissey2007} and converted to a Vega magnitude calculated based on the model atmosphere of Vega from \citet{Kurucz1993}.  The NUV photometry is listed in Table \ref{phot}.  CXOJ1426+35 was not detected by the far-ultraviolet (FUV) detector on \emph{GALEX}.     

\subsubsection{Optical/Near-IR Imaging}
\label{optical_nir}

\begin{figure*}
\centerline{\epsfig{figure=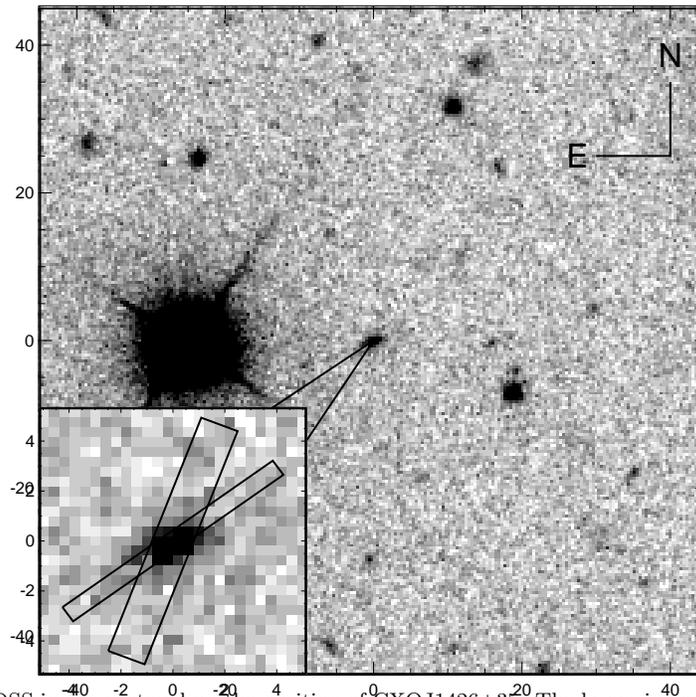,height=3.5in,width=3.5in}}
\caption{{Stacked $g+r+i$ SDSS image centered on the position of CXOJ1426+35.  The larger image is 90$^{\prime \prime}$ on a side with North up and East to the left.  The inset is a close-up of CXOJ1426+35, $10^{\prime \prime}$ on a side, showing that the galaxy is clearly extended with a position angle of P.A.$=-54.1^{\circ}$.  Boxes show the spectroscopic slit configurations obtained using LRIS (P.A.$=-21^{\circ}$, slit width = 1.5$^{\prime \prime}$) and NIRSPEC (P.A.$=-54.1^{\circ}$, slit width = 0.7$^{\prime \prime}$).  Note the bright star, $r=13.04$, $\sim25^{\prime \prime}$ to the east of CXOJ1426+35, making this galaxy an ideal AO target.}}
\label{inset}
\end{figure*}

CXOJ1426+35 is well-detected by the SDSS (Figure \ref{inset}).  In Table \ref{phot} we list the model magnitudes, calculated by the SDSS pipeline using the best fitting function (either a deVaucouleurs or an exponential profile - in this case an exponential profile) in the \emph{r} band and applying that model to the other bands.  We converted the SDSS magnitudes to AB magnitudes following \citet{Bohlin2001}: AB$_{\rm offset}\approx 0.04, 0.0, 0.0, 0.0$, and $-0.02$ magnitudes for $u$, $g$, $r$, $i$, and $z$, respectively, where $M_{\rm AB}=M_{\rm SDSS}-$AB$_{\rm offset}$.  We then converted to Vega magnitudes based on the model atmosphere of Vega from \citet{Kurucz1993}.  Figure \ref{inset} shows the stacked $g+r+i$ SDSS image of CXOJ1426+35, revealing a clearly extended host galaxy. 

Note that the 8.5 deg$^2$ Bo\"otes field has been deeply imaged at optical wavelengths by the NOAO Deep, Wide-Field Survey \citep[NDWFS;][]{Jannuzi2000}.  However, the bright star $\sim25^{\prime \prime}$ East of our target (see Figure \ref{inset}), while useful for the AO observations discussed in Section \ref{ao}, severely contaminates the NDWFS photometry for CXOJ1426+35 due to a bright diffraction spike.

The Bo\"otes field has also been deeply imaged at NIR wavelengths ($J$, $H$, and $K_s$) using the NOAO Extremely Wide-Field Infrared Imager \citep[NEWFIRM;][]{Probst:2004,Probst:2008} on the 4-m Mayall Telescope at Kitt Peak National Observatory.   Those data were obtained as part of the Infrared Bo\"otes Imaging Survey, an NOAO Large Survey Program (Gonzalez et al., in prep.).  CXOJ1426+35 is detected in all bands.  Table \ref{phot} lists the corresponding photometry in both Vega and AB magnitudes, where the latter assume the Vega-AB offsets from \citet{Blanton2005}.

\subsubsection{Mid-IR Imaging}
\label{mid-ir}

CXOJ1426+35 has mid-IR data available from the \emph{Spitzer} Deep, Wide-Field Survey \citep[SDWFS;][]{ashby2009}, a four-epoch infrared survey of 10 deg$^{2}$ in the Bo\"{o}tes field using the Infrared Array Camera \citep[IRAC;][]{Fazio2004} on the \emph{Spitzer Space Telescope}.  Magnitudes are provided for channels 1, 2, 3 and 4, which correspond to bandpasses with effective wavelengths of 3.6, 4.5, 5.8 and 8.0 $\micron$, respectively.  The photometry listed in Table \ref{phot} assumes the position measured in the 4.5 $\micron$ image, and uses 4.0$^{\prime \prime}$ radius aperture photometry, corrected to total magnitudes, and converted from Vega magnitudes to AB magnitudes using constants listed in the IRAC Data Handbook.  The mid-IR colors are AGN-like, falling within the empirical separation of active galaxies from Galactic stars and normal galaxies \citep{Stern2005}, though it does lie on the red end of the overall AGN distribution.

\subsubsection{Radio Imaging}
\label{radio}

CXOJ1426+35 is undetected in the deep, 1.4 GHz observation of the Bo\"{o}tes field by the Westerbork Synthesis Radio Telescope \citep{deVries02}.  These data reach a median noise level of 140$\mu$Jy ($5\sigma$), and at this depth, no sources  are detected within 14.6$^{\prime}$ of CXOJ1426+35.  For a source at a redshift of $z=1.175$, this detection limit corresponds to a radio power of $P_{\rm{1.4~GHz}}=1\times 10^{23}$ W Hz$^{-1}$.  For comparison, we used the 1.4 GHz radio powers of the 14 compact radio sources in \citet{Holt2008} (calculated using the published power law indices) as those sources also exhibit narrow emission lines with velocity offsets, presumably as a result of interactions between the radio jets and the NLR clouds.  The range of radio powers is $P_{\rm{1.4~GHz}}=3.8\times 10^{26} - 1.5\times 10^{28}$ W Hz$^{-1}$, with a median of $P_{\rm{1.4~GHz}}=3.8\times 10^{27}$ W Hz$^{-1}$, meaning this sample of compact radio sources would have all been detected in this observation.  Therefore, CXOJ1426+35 is unlikely to be a similarly strong, compact radio source.

\subsection{New Data and Analysis}

While targeting a mid-IR transient which proved to be a self-obscured supernova at $z \sim 0.2$ \citep{Kozlowski2010}, we obtained a deep spectrum of CXOJ1426+35 at Keck Observatory in March 2010.  As described below, this intriguing spectrum triggered several additional observations.
~\newline
\subsubsection{Keck/LRIS: Optical Spectroscopy}
\label{lris}

\begin{figure*}
\centerline{\epsfig{figure=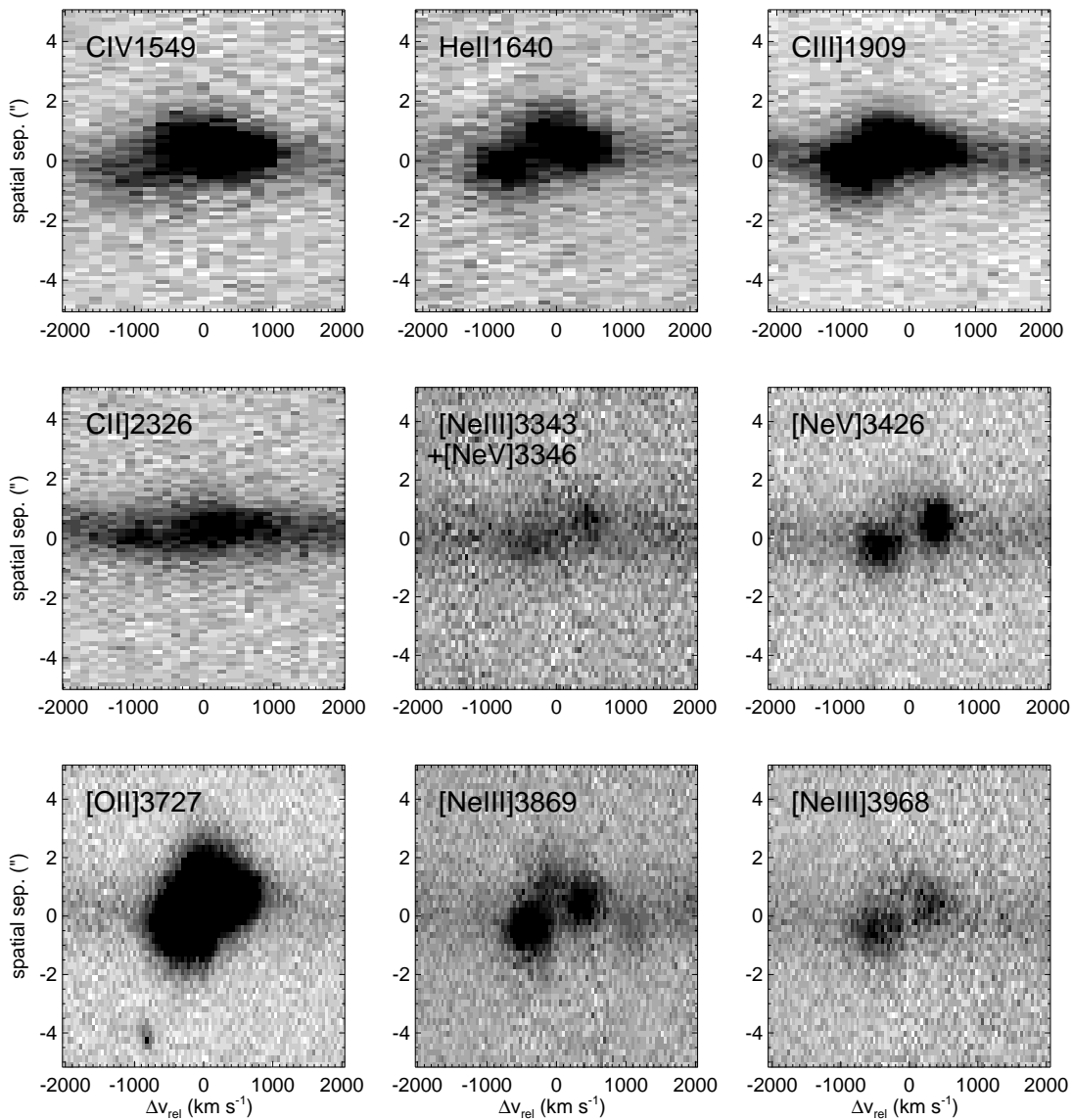,height=6.5in,width=6.06in}}
\caption{{Prominent emission lines seen in the Keck/LRIS 2-D spectrum of CXOJ1426+35.  The two components are most clearly seen in the forbidden lines which are narrower and less blended.  For lines with blended doublets, particularly CIV1549 for which the `blue' component is much weaker than the `red', the two components are more difficult to resolve but still present.  From left to right, the first four panels show emission lines from the blue side of the dichroic, and the last five panels are from the red side of the dichroic.  The dispersion scale is 1.90 $\rm \AA$ pixel$^{-1}$ for the blue side and 1.18 $\rm \AA$  pixel$^{-1}$ for the red side.  Note that the blue side was binned by two pixels in the dispersion direction.  Along the spatial axis the scale is 0.135 arcsec pixel$^{-1}$.} In each box the horizontal axis (relative velocity) spans 4000 km s$^{-1}$ and the vertical axis (spatial separation along the slit) spans 10$^{\prime \prime}$.  At a P.A. of $-21^{\circ}$, the spatial separation between the two components is $0.48^{\prime \prime}$ and the velocity separation is 790 km s$^{-1}$, as measured from [NeV]3426.  \\  }
\label{lris2d}
\end{figure*}

\begin{figure*}
\centerline{\epsfig{figure=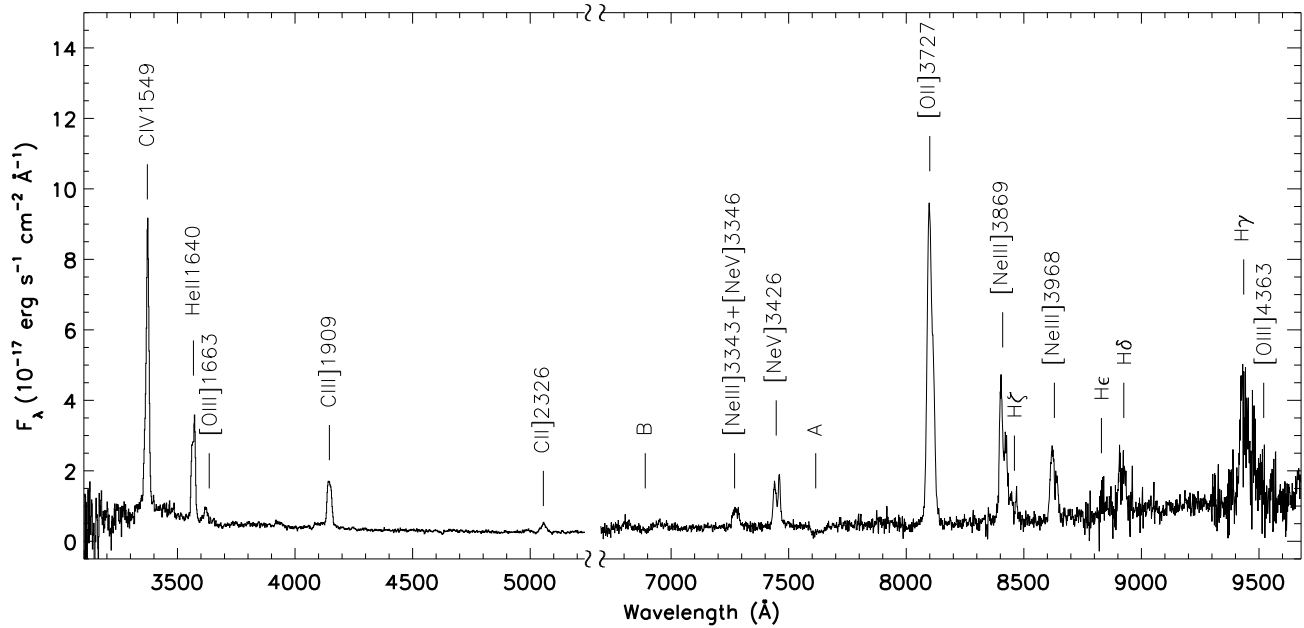,height=3.4in,width=7in}}
\caption{{Keck/LRIS spectrum of CXOJ1426+35, in the observed frame, extracted from a 3$^{\prime \prime}$ diameter region including both velocity components, mimicking the source spectrum through an SDSS fiber.  The emission lines and atmospheric absorption lines (A and B band) are labeled.  Note the double high ionization narrow emission lines, suggestive of two Type 2 AGN in the process of merging.  Two dimensional spectra of these emission lines are presented in Figure \ref{lris2d}.}}
\label{lris_b+r}
\end{figure*}

Optical spectra of CXOJ1426+35 were obtained with the Low Resolution Imaging Spectrometer \citep[LRIS;][]{Oke1995} on UT 2010 March 12 using the D680 dichroic and the 400/3400 grism on the blue side and the 400/8500 grating on the red side.  The 1.5$^{\prime \prime}$ slit was oriented at P.A.$=-21^{\circ}$.  The data consisted of three exposures of 1200 seconds each.  For a 1$^{\prime \prime}$ slit, this instrument configuration provides a resolution of $6.5-7.1\rm~\AA$ (FWHM) on the blue side of the dichroic and 6.9$\rm~\AA$ (FWHM) on the red side of the dichroic for objects filling the slit.  We processed the data using standard techniques within IRAF.  The 2D-spectra (Figure \ref{lris2d}) show two emission line systems separated both spatially and along the dispersion axis.  At a P.A. of $-21^{\circ}$, the spatial separation is $0.48^{\prime \prime}$, as measured from the [NeV]3426 line.  To model the emission line velocity-offsets and widths, we extracted an aperture of $3^{\prime \prime}$ in diameter to recover the flux from the entire source (both emission line systems plus the continuum), and separate apertures centered on each component to isolate the flux from the `blue' and `red' systems.  Figure \ref{lris_b+r} shows the {\bf $3^{\prime \prime}$} extraction with the emission line identifications.  These extractions mimic what a SDSS fiber spectrum of the source would look like, though the spatially resolved spectrum, shown in Figure \ref{lris2d}, provides considerable additional information.

The extracted spectra were corrected for Galactic extinction using a color excess of $E(B-V)=0.015$ determined from the model of \citet{Schlegel98}.  We used SPECFIT \citep{Kriss94} to model each emission line as two Gaussian profiles (for the `blue' and `red' emission line systems, respectively) on top of a local continuum.  Unfortunately, the individual component apertures have significant contamination from each other, making it difficult to deconstruct the emission line flux attributable to each emission line system.  Therefore, those apertures were used only to obtain estimates for the observed central wavelengths of each emission line, and the lines were fit again in the $3^{\prime \prime}$ aperture to determine the spectral model parameters and associated errors.  For the highly blended emission lines, some of the Gaussian parameters were fixed to those of less blended lines with similar ionization potentials.  Doublets with rest wavelengths close together were modeled as single Gaussians. We also detected an absorption feature which is consistent with an H9 absorption line at a redshift intermediate to the two emission line systems.  For each free parameter that was fit, the quoted errors correspond to $1\sigma$ uncertainties, and these errors were propagated into any further calculations.  The line fluxes, FWHMs, redshifts and velocities separations are presented in Table \ref{lines}. 

For both the `blue' and `red' systems each emission line is well-fit by a single Gaussian.  As measured from [NeV]3426, the redshifts for the `blue' and `red' systems are $z=1.1715$ and $z=1.1773$, respectively. The range of spectral coverage is $3100-5200\rm~\AA$ on the blue side of the dichroic and $6700-9600\rm~\AA$ on the red side of the dichroic, which corresponds to rest-frame wavelength ranges of $1430-2380\rm~\AA$ and $3080-4410\rm~\AA$ and allows us access to a range of rest-frame optical and UV emission lines that are useful in diagnosing the structure and kinematics of the NLR gas from their velocities and line widths.  If the absorption line is interpreted as H9 absorption within the host galaxy, this yields a systemic redshift of $z=1.1751$.  At this redshift, the emission line velocity-splitting is 790 km s$^{-1}$ as measured from [NeV]3426.  With no broad components detected in any of the optical/UV emission lines, the spectrum of CXOJ1426+35 resembles that of a Type 2 AGN, which in the unified theory of AGN implies that the BLR and central continuum source are obscured.  This is consistent with the heavy obscuration implied by the X-ray data.

\subsubsection{Keck/NIRC2: Adaptive Optics Imaging}
\label{ao}

\begin{figure*}
\centerline{\epsfig{figure=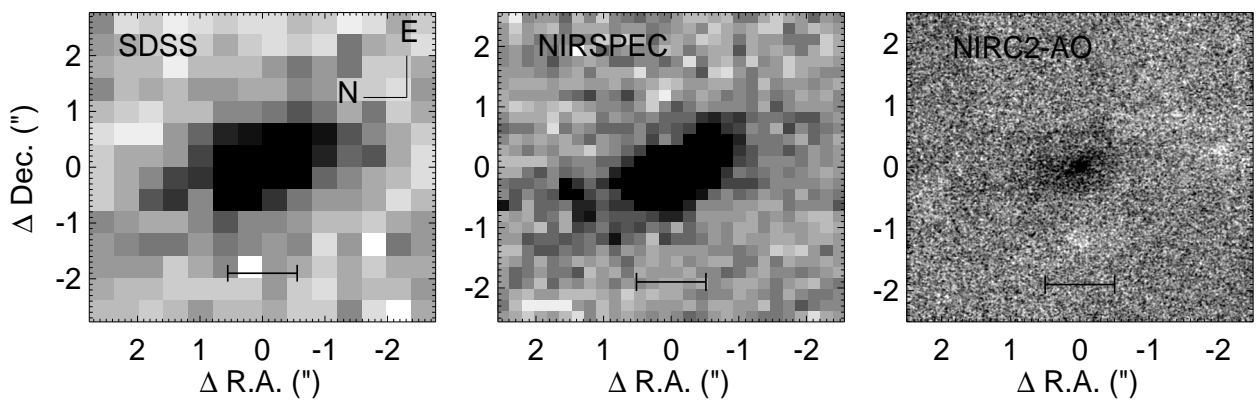,height=2.12in,width=7in}}
\caption{{From left to right: stacked $g+r+i$ SDSS (0.40 arcsec pixel$^{-1}$), Y-band NIRSPEC (0.18 arcsec pixel$^{-1}$) and $K^{'}$-band NIRC2 - AO (0.01 arcsec pixel$^{-1}$) images.  In each image, the scale bar represents 10 kpc at $z = 1.175$.  The AO image has been smoothed by convolving with a Gaussian function using the `gauss' task in IRAF.}}
\label{multi}
\end{figure*}

Adaptive optics (AO) imaging of CXOJ1426+35 was obtained by the narrow camera of the Near Infrared Camera 2 \citep[NIRC2;][]{vanDam2004} at the Keck II telescope on UT 2010 June 30.  These data consisted of three 180 second data exposures at P.A.$=0.0^{\circ}$ through the $K^{'}$ filter with an effective wavelength of $\lambda_{\rm eff}=2.12~\micron$ which corresponds to a rest-frame wavelength of $\lambda_{\rm eff}\sim1\micron$.  An average dark field image was created using `imcombine' and was subtracted from each data image.  Each image was flat-fielded and then differenced from the other science frames.  The images were shifted using the position of the target and then stacked using `imcombine'.  The final stacked image, 10$^{\prime \prime}$ on a side, is shown in Figure \ref{multi} and compared to the $g+r+i$ SDSS image and the NIRSPEC image (see Section \ref{sec:nirspec}).  The image reveals a single, diffuse source.  The lack of two nuclei in the $K^{'}$ band implies that either there is no second AGN or alternatively that there is significant obscuration around $1\micron$.  Significant obscuration would also be consistent with the high column density implied by modeling of the X-ray data (Section {\ref{Chandra}}) and with the extinction implied by our modeling of the spectral energy distribution (Section \ref{sed_model}).           

\subsubsection{Keck/NIRSPEC: Near-IR Imaging and Spectroscopy}
\label{sec:nirspec}

NIR imaging and spectroscopy of CXOJ1426+35 was obtained using the Near Infrared Spectrometer \citep[NIRSPEC;][]{McLean1998} at the Keck II telescope on UT 2010 July 18.
 
\emph{Imaging:} The imaging consisted of 11 dithered images of 10 seconds each.  The filter used was the NIRSPEC-1 filter which has a wavelength range of $0.95-1.12~\micron$ and is approximately a Y-band filter.  The images were flat-fielded and then differenced.  The background was subtracted using `background' for each quadrant of each image independently and cosmic rays were removed using `xzap' in IRAF.  The images were stacked with `imcombine' using for offsets the brightest star, when available, and the target otherwise.  At a redshift of $z=1.175$, the filter provides rest-frame coverage of $0.44-0.52~\micron$.  The final, combined image is presented in Figure \ref{multi} and reveals that the host galaxy dominates at optical wavelengths and consists of an elongated and ``lumpy" structure that hints at a disturbed morphology.

\emph{Spectroscopy:} We obtained NIRSPEC spectroscopy in low-resolution mode using a 0.7$^{\prime \prime}$ wide slit placed at a position angle of P.A.$=-54.1^{\circ}$ so that the spatial axis was parallel to the major axis measured in the SDSS image.  For a $0.38^{\prime \prime}$ slit, the spectrograph provides a resolving power of $\sim2,200$.  At two positions, A and B, along the slit, four 300 second exposures were taken in the order ABBA.  The same filter as for imaging, NIRSPEC-1, was used.  The spectra were differenced, cleaned of cosmic rays using the IRAF task `szap' and background subtracted using `background'.  The spectra were shifted and stacked using `imcombine'.  

\begin{figure*}
\centerline{\epsfig{figure=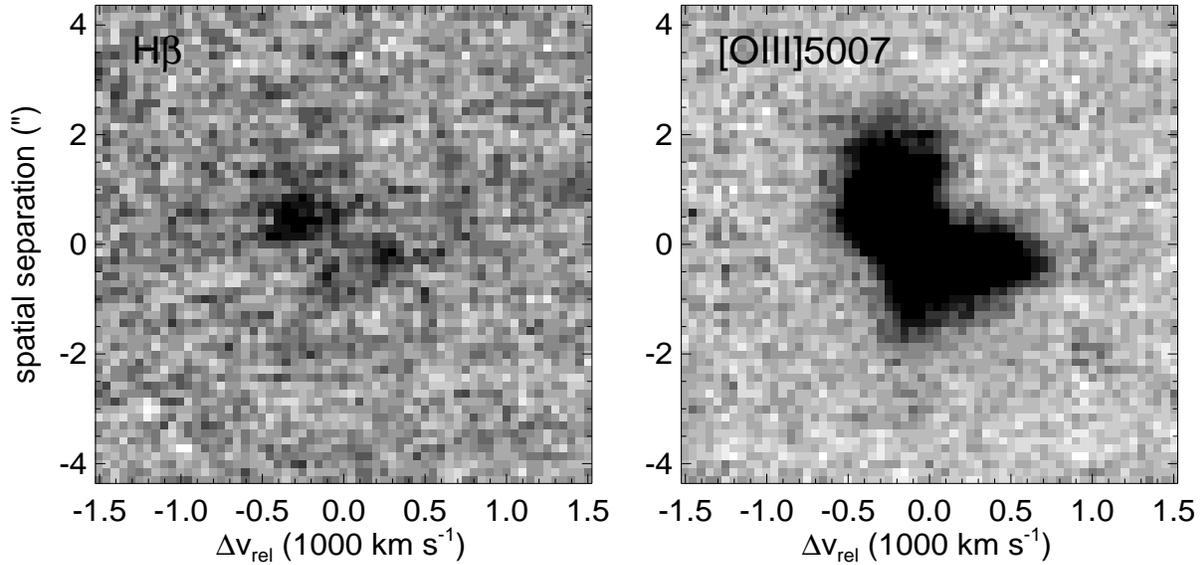,height=3.1in,width=6.5in}}
\caption{{Panel showing the H$\beta$ and [OIII]5007 emission lines in the 2-D NIRSPEC spectrum taken at P.A. of -54.1$^{\circ}$.  The dispersion scale is 2.10$\rm~\AA$ pixel$^{-1}$, and the scale along the spatial axis is 0.143 arcsec pixel$^{-1}$.  In each box the horizontal axis (relative velocity) spans 3000 km s$^{-1}$ and the vertical axis (spatial separation along the slit) spans $9^{\prime \prime}$.  At P.A. of -54.1$^{\circ}$ (aligned parallel to the major axis of the galaxy), the spatial separation between the two components is $0.69^{\prime \prime}$ and the velocity separation is 690 km s$^{-1}$, as measured from [OIII]5007.} }
\label{nirspec2d}
\end{figure*}

The range of spectral coverage allows us access to the H$\beta$, [OIII]4959 and [OIII]5007 lines, which clearly show two emission line components (Figure \ref{nirspec2d}).  As measured from [OIII]5007, the separation between the two components along the spatial axis (parallel to the major axis of the galaxy in the NIRSPEC image) is $0.69^{\prime \prime}$.  Aperture extractions and spectral modeling proceeded as with the LRIS spectrum.  Three apertures were extracted that include a region of 3$^{\prime \prime}$ diameter encompassing the entire source, and separate apertures for the each of the `blue' and `red' systems.  The extracted spectrum was corrected for Galactic extinction, and spectral modeling was performed on the $3^{\prime \prime}$ aperture with `SPECFIT' with initial guesses for the line central wavelengths determined from the individual apertures for each component.  The `red' H$\beta$ component is very weak and therefore we have fixed its FWHM to that of the `red' [OIII]5007 component and adopted the measured flux as an upper limit.  Line fluxes, FWHMs, redshifts and emission line velocity-splittings are listed in Table \ref{lines}.

\begin{figure*}
\centerline{\epsfig{figure=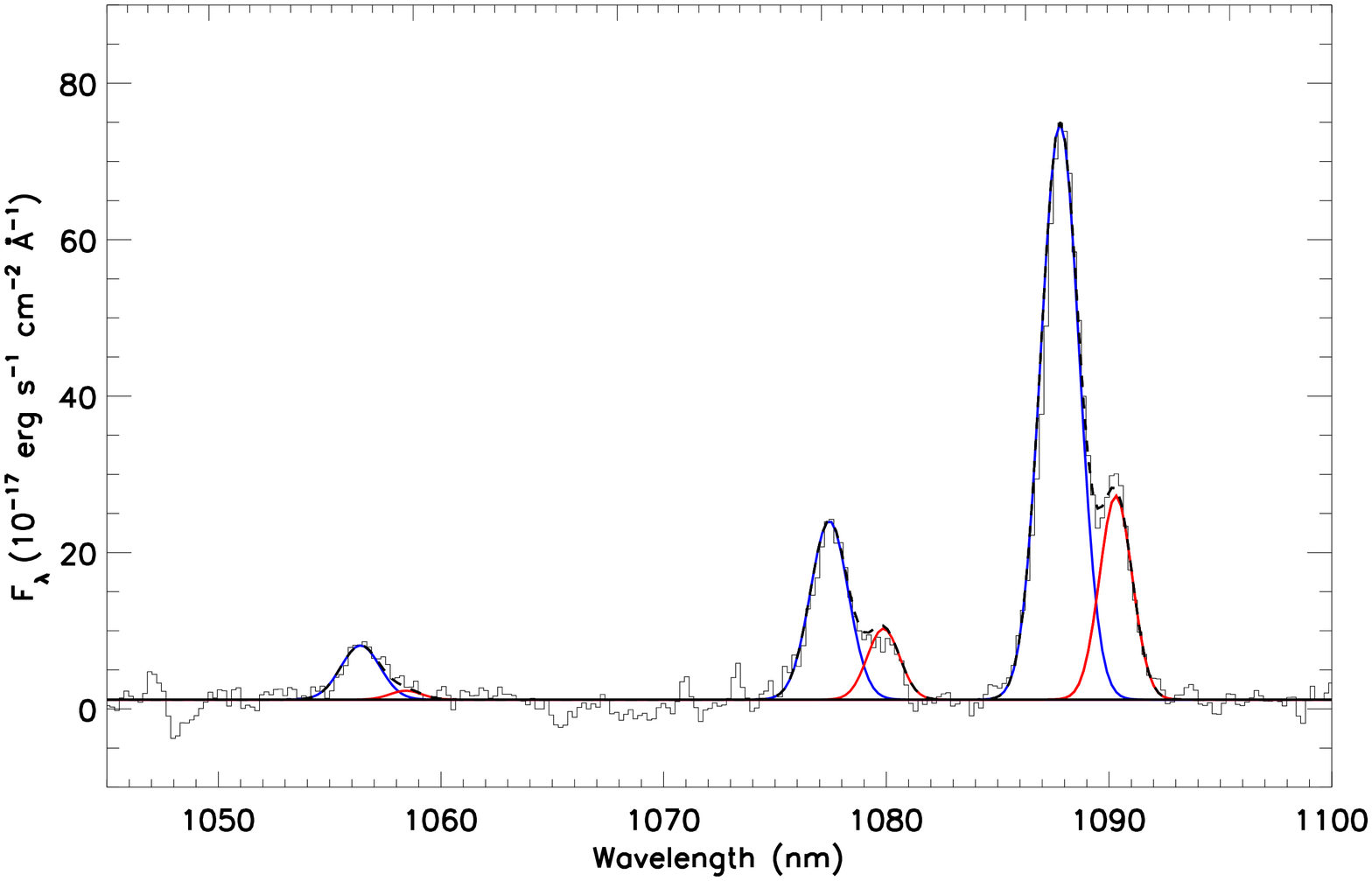,height=4.7in,width=6.5in}}
\caption{{NIRSPEC spectrum of CXOJ1426+35, in the observed frame, extracted from an aperture 3$^{\prime \prime}$ in diameter that includes both velocity components.  The emission lines are H$\beta$, [OIII]4959 and [OIII]5007.  Overlaid on the spectrum are the best-fit Gaussians.  Each emission line is fit by two Gaussians, one for the `blue' and one for the `red' emission line component.}}
\label{nirspec}
\end{figure*}

The emission lines for both the `blue' and `red' systems are each fit by a single Gaussian (Figure \ref{nirspec}). The [OIII]5007 line has the highest S/N and the wavelength solutions for the Gaussian peaks provide a line of sight velocity separation of 690 km s$^{-1}$.  The ratios between H$\beta$ and H$\gamma$ are consistent within their errors with case B recombination theory for a reasonable NLR temperature range of 2,500-20,000 K \citep{Osterbrock:2006}.  Considering the significant error of the H$\gamma$ flux measurement, we do not find any measurable Balmer decrement in the narrow emission line systems. The ratio between [OIII]5007 and H$\beta$ is highly useful in diagnosing the nature of the ionizing continuum  through its placement on a BPT diagram \citep{Baldwin1981}.  The line fluxes provide $\rm [OIII]5007/H\beta$ ratios of $8.9\pm 3.2$ for the `blue' system and a lower limit of $11.5$ for the `red' system.  Though we do not have access to H$\alpha$ and can not uniquely place the emission line systems on a typical diagnostic diagram \citep{Kauffmann:2003,Kewley:2006}, the [OIII]5007/H$\beta$ ratios put both systems above the demarcation defined in \citet{Kewley:2001}, clearly indicative of AGN and consistent with the high ionization lines observed in the optical spectrum.

\subsection{Modeling of the Spectral Energy Distribution}
\label{sed_model}

\begin{figure*}
 \centerline{\epsfig{figure=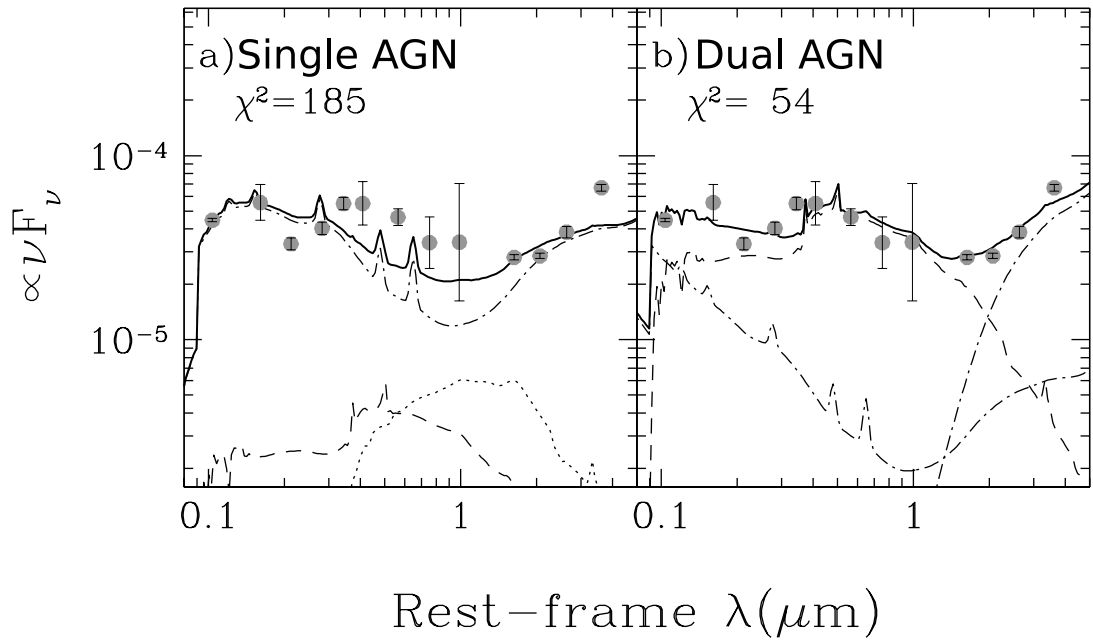,scale=0.75}}
  \caption{\footnotesize{SED fits to the broad-band photometry assuming a single
    ({\textit{panel a}}) and dual ({\textit{panel b}}) AGN components. The best-fit AGN components in each case are shown by the dot-dashed lines, while the Im component is shown by the dashed lines and the Sbc component is shown by the dotted lines \citep[see][for details on the SED templates]{assef2010}. The solid line shows the best-fit SED, which corresponds to the addition of the different components.}}
  \label{sed_vfv}
\end{figure*}

Motivated by the high level of AGN obscuration implied from the X-ray HR and by the SDSS and NIRSPEC images, which suggest that the galaxy dominates at optical wavelengths, we have attempted to separate the SED into galaxy and AGN components to determine their relative contributions.  This was done to determine if the AGN extinction inferred from the SED is consistent with the X-ray HR \citep{Hickox:2007} and, furthermore, if the SED is potentially consistent with two AGN.  We modeled the broad-band photometry of \obj\ using the SED templates for AGNs and galaxies of \citet{assef2010}. The four empirically constructed SED templates that form this basis extend from 0.03 to 30$~\mu$m and include an old stellar population (referred to as E), an intermediate star forming population (Sbc), a strongly star-forming population (Im), and an AGN component. We refer the reader to \citet{assef2010} for details on the SED templates. The standard approach of \citet{assef2010} is to fit the broad-band photometry with a non-negative linear combination of these four templates, allowing for additional dust extinction for the AGN component (and only the AGN component).  However, this approach does not allow for a satisfactory fit to the broad-band photometry ($\chi^2 = 185$; Fig. \ref{sed_vfv}a). Led by the possibility of \obj\ being a dual AGN, we have modified our analysis and attempted to model the SED with two independent AGN components, each with a different amount of extinction, in addition to the host components. This approach yields a much better fit ($\chi^2=54$), as shown in Figure \ref{sed_vfv}b. An unreddened low luminosity AGN component is used to model the UV excess over the galaxy-host SED, while another AGN component, intrinsically 10 times more luminous and with large amounts of obscuration ($E(B-V)=3.44$), is used to model the mid-IR emission. This fit, while much better than that obtained with the standard approach of \citet{assef2010}, is still not completely satisfactory; the $i-$band, which includes redshifted [OII]3727 emission, is underpredicted by $\sim3.5~\sigma$, and the IRAC $8.0\micron$ band is underpredicted by $\sim4~\sigma$. This does, however, lend credibility to the hypothesis that \obj\ is a dual AGN. The SED modeling is compatible with the X-ray HR - the color excess implies $n_{H}\approx 2\times 10^{23}$ cm$^{-2}$, assuming an SMC-like gas/dust ratio \citep{Maiolino:2001} - and with the single source detected in the AO image, as the more luminous but highly reddened AGN component would dominate the X-ray emission but would be virtually invisible at rest-frame $1~\mu$m ($A_{1\micron}=4.66$ mag from the extinction law of \citep{Cardelli:1989} and for $R_{V}=3.1$.). The optical broad lines of the less luminous, unreddened AGN (with $\sim10\%$ of the host luminosity) would be too faint to be detectable \citep{Hopkins:2009}, consistent with our observations.  We also find no significant evidence for a broad component in the UV lines, though in this model a broad component in CIV1549 would possibly be detectable with sufficient signal quality.  The relative luminosities of the two AGN components in the dual AGN model are consistent within the factor of 4 uncertainty of the bolometric luminosities, $5\times 10^{46}$ erg s$^{-1}$ and $1.5\times 10^{46}$ erg s$^{-1}$ for the `blue' and `red' components, respectively, from the $L_{\rm [OIII]}$ bolometric correction for Type 2 AGN in \citet{Lamastra:2009}.

We note that an alternative approach would be to associate the UV excess with a very young stellar population instead of a second, lower luminosity AGN.  Such a scenario would imply a star-formation rate of approximately 40$~M_{\odot}~\rm yr^{-1}$ using the total UV luminosity at $2150\rm~\AA$ and the relation of \citet{Madau:1998} for a Salpeter IMF, and a stellar mass of approximately $10^{10}~M_{\odot}$ using the $g-r$ color and $M/L_{K}$ relation of \citet{Bell:2003} for the host. A specific star-formation rate of $\sim 4\times10^{-9}~\rm yr^{-1}$ is close to the high end of the distribution at the redshift of CXOJ1426+35 \citep[see][]{Cooper:2008} but it is not implausibly large, making this modeling of the SED a viable alternative to the dual AGN scenario.

\section{Interpretation}
\label{interpretation}

In this section, we discuss six physical scenarios which may account for the unusual double emission line peaks observed in the restframe optical/UV lines, and the morphology as evident from the images, particularly the lack of a double nucleus in the NIRC2-AO image.  These scenarios are a chance superposition, a rotating accretion disk, jet/cloud interactions, an unusual NLR geometry, accretion disk outflows, and a dual AGN.  Similar explanations have been invoked for other dual AGN candidates and sources with double-peaked or offset emission lines (in particular, see the discussions of \citealp{Gerke2007}, \citealp{Komossa2008b}, and \citealp{XK09}), though we elaborate upon them here.  We discuss the applicability of each scenario to CXOJ1426+35, concluding that a dual AGN is a viable explanation.  

\subsection{Superposition}

A chance superposition of two unrelated sources has occasionally been posited as a possible interpretation of the multiple narrow emission lines seen in dual AGN candidates.  Ignoring the fact that, in this case, the interpretation is rather forced since two systems with this little separation both physically and in velocity space are unlikely to be unrelated, we can use the surface density of AGN to estimate the probability of such a chance superposition.  Using the SDSS quasar catalogue \citep{Schneider07} to determine the surface density of luminous (brighter than $M_{i}=-22.0$) quasars at $1.165 < z < 1.185$ (e.g., a redshift bin equivalent to nearly one hundred times the velocity difference between the two systems), we find a probability of $\sim 3 \times 10^{-8}$ that CXOJ1426+35 is due to a chance superposition of field galaxies.    Though the SDSS quasars are luminous Type 1 quasars and do not include obscured sources such as CXOJ1426+35, this small probability strongly argues against a superposition.  However, the probability would increase were CXOJ1426+35 a member of a cluster \citep[e.g.][]{Dotti:2010}.  Using the cluster surface density at the high end of the surface density distribution in \citet{Dressler:1997}, we find that the probability increases by a factor of $\sim15$, implying that even in a cluster this scenario is highly improbable.

\subsection{Jet/Cloud Interaction}

Radio jets produced by the central active SMBH are capable of introducing a variety of kinematics into the NLR.  Well-collimated jets can produce asymmetric and/or blueshifted narrow emission line profiles, and indeed jet/cloud interactions in the inner NLR have been used to explain the [OIII] blueshifts seen in some compact radio sources \citep{Holt2003,Holt2008,Fu2009}.  If the redshifted jet is not obscured, this scenario can explain the double-peaked narrow emission lines of CXOJ1426+35, and this was shown to be the case for a previous dual AGN candidate \citep{Rosario:2010}.  Nevertheless, some questions remain as to the capability of radio plasma to entrain the bulk of the NLR so that the velocity-shifted lines dominate in the emergent spectrum, rather than introduce instabilities that fragment the clouds.  For example, \citet{XK09} discuss for another dual AGN candidate the scenario in which only a single cloud interacts locally with each side of the jet, but find that the [OIII]5007 line luminosity is not consistent with just two clouds, and we arrive at the same conclusion for CXOJ1426+35 given its high [OIII]5007 luminosity.  

However, an even tighter constraint on the possibility of a jet/cloud interaction in CXOJ1426+35 is derived from the radio power corresponding to the radio detection limit (Section \ref{radio}).  The sample of radio sources in \citet{Holt2008} would have all been well-detected.  Though these are only 14 sources and do not represent all radio galaxies, they are useful for comparison as they show blueshifted [OIII]5007 lines, presumably a result of local jet/cloud interactions in the inner NLR.  If jet/cloud interactions are responsible for producing the double-peaked narrow emission lines, then CXOJ1426+35 is likely to be a strong radio source, similar to the radio sources in \citet{Holt2008}, and would therefore have also been detected.  However, the lack of a detection makes the presence of (at least strong) jet/cloud interactions highly unlikely.

\subsection{Rotating Accretion Disk}
\label{rotating}

A rotating disk with a line emitting surface is capable of producing double-peaked emission.  The relative positions of the peaks and the shape of the lines are dependent on geometrical factors which can enhance the effect, such as the orientation of the disk plane relative to the observer's line of sight, and the disk outer radius.  Since this disk model has been successfully applied to accretion disks around SMBHs, providing a possible explanation for the asymmetric (non-Gaussian), broad and often double-peaked Balmer emission lines of some AGN and quasars, including some candidate binary SMBHs (see Section \ref{intro}), we consider this scenario for CXOJ1426+35 but find that it is unlikely: the 2-D spectroscopy reveals that the line peaks are at a projected separation of $\sim5.5$ kpc.  This is much larger than the sub-pc sizes of such disks based on SMBH masses and H$\alpha$ line profiles \citep{CH89} and makes the single accretion disk interpretation of CXOJ1426+35 unphysical.

\subsection{Unusual NLR Geometry}
\label{NLRGeo}

One may imagine a disk-shaped NLR which can, in principle, produce the double-peaked lines in a way analogous to the rotating disk in Section \ref{rotating}.  One advantage of this scenario is that NLRs are much larger than accretion disks and could explain the observed separation.  While a size of 5.5 kpc is large for a typical Seyfert galaxy \citep{Schmitt:1996}, luminous quasars have been observed with NLRs of up to 10 kpc \citep{Bennert:2002}.  The total [OIII]5007 luminosity of CXOJ1426+35 is nearly an order of magnitude larger than that of the brightest quasar in the sample of \citet{Bennert:2002}, suggesting that the NLR of CXOJ1426+35 may plausibly extend to the observed separation of 5.5 kpc.

However, an additional and important consideration involves line ratios and line strengths: in this scenario both line systems should have similar line ratios and line strengths since the ionizing continuum incident on the gas and gas densities would be the same since they would be part of the same NLR.  This scenario was found to be plausible for a subset of SDSS double-peaked AGN with even [OIII] flux ratios in \citet{Smith:2011}.   However, from Table \ref{lines}, it is clear that the line ratios and line strengths, particularly for CIV1549 and [OIII]5007, are not similar.  Unless one side of the NLR suffers more extinction than the other, and there is no reason to expect this, the different line strengths are inconsistent with both line systems originating from the same NLR.   

\subsection{ Accretion Disk Outflows}
\label{outflows}

Winds moving in opposite directions and ionized by a central continuum source can, in theory, produce the double-peaked profile of the narrow emission lines (see the discussions in \citealt{XK09} and \citealt{Fischer:2011}).  Mechanisms for producing such an outflow include radiation pressure acting on gas and dust \citep{Everett2007b} or a hot wind that entrains the NLR clouds \citep{Everett2007a}.  A potential problem with this scenario is the difficulty of the winds extending to typical NLR scales in order to affect the whole NLR.  However, a SMBH accreting at a relatively high Eddington ratio ($f_{\rm Edd}=L_{\rm BOL}/L_{\rm Edd}$) may be capable of producing a wind strong enough to extend to kpc scales.  In general, luminous, Type 1 unobscured quasars at $z>0.5$ may have typical Eddington ratios of $f_{\rm Edd}\approx 0.25$ \citep{Kollmeier2006}, though recent work by \citet{Kelly:2010} suggests that $f_{\rm Edd}$ peaks near 0.05, or alternatively AGN might instead accrete close to Eddington in a `buried' phase as described in \citet{King:2010}.  In the sample of nine narrow line Seyfert 1 galaxies (NLS1s) with [OIII] offsets in \citet{Komossa2008b} the Eddington ratios are high ($f_{\rm Edd}=0.5-1.5$, with the average value $f_{\rm Edd}=0.9$), and 
the authors remark that these high Eddington ratios may capable of producing the strong outflows needed to explain the observed velocity shifts.  In principle, a very rough estimate of the Eddington ratio can be made using the bolometric luminosity estimated in Section \ref{Chandra} and estimating a SMBH mass assuming all of the stellar mass (calculated in Section \ref{sed_model}) is within the bulge and using the $M_{\rm BH}\sim M_{\rm Bulge}$ relation \citep{Haring:2004}.  However, the resulting mass of $1.2\times 10^{7}$ M$_{\odot}$, and the relation $L_{\rm Edd}=1.3\times 10^{38}$ M$_{\rm BH}$/M$_{\odot}$ imply that the bolometric luminosity is $\sim50$ times that of the Eddington limit.  This result seems unphysical, possibly because this galaxy is still in the process of forming and the $M_{\rm BH}\sim M_{\rm Bulge}$ relation has yet to be established in this system.  However, a different test of the outflow scenario uses the optical/UV emission lines as diagnostic indicators of the ionization parameters and kinematics of the `blue' and `red' systems.  We perform two calculations based upon the emission line parameters that provide clues as to the origin of the double-peaked lines.

\emph{Ionizing Continuum:} We can use the ionizing continuum to estimate the maximum distance at which the emission line regions can be so ionized to determine if the outflow scenario is plausible.  The ionizing continuum is described by $Q[H^{0}]$, which we have estimated from the total H$\beta$ luminosity of both the `blue' and `red' sources via the relation $L_{H\beta}=h\nu_{\rm H\beta}(\alpha^{\rm eff}_{\rm H\beta}/\alpha_{B})(\Omega/4\pi)Q[H^{0}]$.  The expected distance can then be found from the relation $U=Q[H^{0}]/(4\pi r^{2}cn_{H})$ where $U$ is the ionization parameter estimated as a function of gas density from the [OII]3727/[OIII]5007 ratios \citep{Komossa:2006}.  The calculated distances are highly dependent on the global covering factor, $\Omega$, of the winds as viewed from the central source.  Therefore, we adopt a lower limit of $\Omega=1.6$ \citep{Blustin:2009} for each source, which corresponds to an upper limit on the distances.  Assuming a pure Hydrogen nebula and complete ionization ($n_H=n_e$), we estimate the distances for a typical range of NLR electron densities ($n_e=10^{2}-10^{4}$ cm$^{-3}$), where $10^{4}$ cm$^{-3}$ is an upper limit based on the critical density for de-excitation of [OII]3727, which is well-detected in the spectrum of CXOJ1426+35. Given the range in electron densities, we find maximum distances of $2-5$ kpc and $2-6$ kpc for which the `blue' and `red' systems, respectively, can be so ionized.  While these distances are consistent with the projected separation of 5.5 kpc, making the outflow scenario plausible on these grounds, it should be noted that these estimates are intended as upper limits, and that the projected separation is a lower limit on the actual physical separation.  Therefore, these calculations merely show that within the uncertainties of the geometry of the system, two clouds with the observed [OII]3727/[OIII]5007 line ratios and H$\beta$ luminosities could be ionized by a single source.  The limits of this argument are apparent when one considers the significantly different line ratios and line strengths between the two systems as noted in Section \ref{NLRGeo}.  In particular, the significantly different CIV1549 and [OIII]5007 strengths are not expected if the outflowing clouds are at the same distance from the AGN and have similar compositions.  However, in principle, a clumpy obscuring geometry could account for this observation.

\begin{figure*}
\centerline{\epsfig{figure=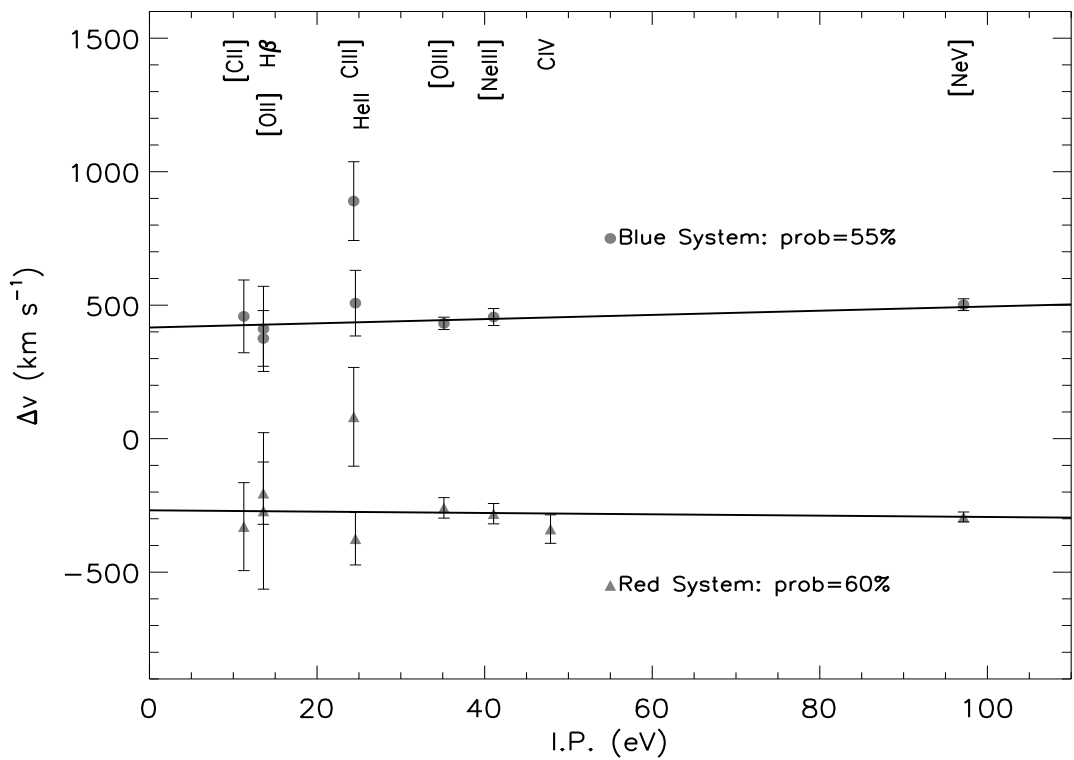,height=4.29in,width=6in}}
\caption{{Plot of emission line velocity-splittings for the `blue' and `red' systems relative to $z=1.1751$ (i.e. the systemic velocity as measured from the H9 absorption line in Section \ref{lris}), where positive velocities indicate a blueshift.  The plotted data points correspond to emission lines for which the positions were allowed to vary during the fit.  Some of the line labels have been shifted for clarity.  The straight lines are the linear least squares best fits.  The quoted probabilities for each system are the two-sided significances, based a Spearman rank correlation, of the deviation from the null hypothesis that there is no evolution of $\Delta V$ with I.P.  In both systems, the probabilities provide no indication of a correlation.  The apparent blueshift of CIII]1909 in each system may be the result of an increase in the $I(\lambda1907)/I(\lambda1909)$ ratio (see Section \ref{outflows}).}}
\label{vsplit}
\end{figure*}

\emph{Ionization Stratification:} The outflow scenario should produce a positive correlation between velocity offset, $\Delta V$, and ionization potential, I.P., otherwise known as an `ionization stratification.'  This is because the high ionization lines will originate closer to the central source where the flux is large enough to produce such lines, whereas low-ionization lines will originate further out where they are less affected by the outflow.  Such a line stratification is evident in the sample of NLS1s with blueshifted [OIII]5007 lines from \citet{Komossa2008b}.  We investigate $\Delta V$ vs I.P. for CXOJ1426+35 using the emission lines for which the central wavelength was allowed to vary during the fitting procedure.  From a Spearman rank correlation test, we find probabilities for a correlation between $\Delta V$ and I.P. of 55\% and 60\% for the `blue' and `red' systems, respectively.  These probabilities indicate that a significant correlation is not apparent given the available line measurements.  In Figure \ref{vsplit} we plot $\Delta V$ vs I.P. for each system along with the linear least squares best fit.  In both systems, the velocities are all consistent within their errors, with the exception of CIII]1909.  Since the CIII]1909 apparent blueshift appears in both systems it may be related to the anomalous CIII]1909 blueshift seen in some quasars as a result of the increased intensity of the $\lambda1907$ transistion relative to $\lambda1909$ ($\sim300$ km s$^{-1}$ offset) at low densities \citep{Ferland:1981}.  While there is no apparent correlation, it should be noted that in the \citet{Komossa2008b} sample, two galaxies have the [FeX] line from the coronal line region (I.P.$=235$ eV), significantly higher than the highest I.P. line observed for CXOJ1426+35 ([NeV], I.P.$=97$ eV).  For the four NLS1s in that sample that do not include the [FeX] line, the trends are comparable to that of CXOJ1426+35.  This indicates there is a possibility that a line stratification might become apparent with a wider range of I.P.s.    

\subsection{Galaxy Merger and Dual SMBHs}
\label{binary}

A system of two SMBHs within the host galaxy, with at least one of them active, would naturally explain the two emission line systems as each SMBH would provide its own NLR.  
Since such a system would arise from a galaxy merger, one may speculate that it is more likely to be found in a disturbed system.  Indeed, the morphology of CXOJ1426+35, particularly the ``lumpy" structure evident in the NIRSPEC image in Figure \ref{multi}, suggests a possible disturbance of some kind.  
This possibility seems even more likely when one considers that, at $z=1.175$, galaxy mergers were more frequent than in the local universe (\citealp{Bridge:2010}).
We also speculate that the X-ray spectrum (which implies a high level of dust extinction) is indicative of a galaxy that has recently undergone a merger.  Therefore, the interpretation of CXOJ1426+35 hosting two SMBHs is quite plausible and worthy of consideration.  We consider two possibilities involving a  galaxy merger and formation of a dual SMBH system.

\emph{One AGN Illuminating Two NLRs:}  This scenario was also suggested for a candidate dual SMBH in  \citet{XK09}.  It is possible that after the merger, only one of the SMBHs would be active but both would have their own NLR.  The active AGN may therefore be able to ionize both NLRs which would have a velocity difference imparted to them by the merger.  For the NLR around the inactive SMBH, assuming it is completely ionized by the active SMBH and with $n_{e}=10^{2}$ cm$^{-3}$, we estimate a free-fall timescale of $\sim5\times 10^{6}$ years.  This is a relatively short time, and it is unlikely that we would see the system during this window.  However, a more rigorous test for this scenario uses the expectation that the ionization parameter and Hydrogen number densities should be consistent with one NLR being further from the ionizing source than the other.  In Section \ref{outflows}, we estimated a range of distances based on the ionization parameters (estimated from the [OII]3727/[OIII]5007 ratio) and a typical range of electron densities ($n_e=10^{2}-10^{4}$ cm$^{-3}$).  It is theoretically possible to obtain an independent estimate for the electron densities from the flux ratios of emission lines produced by different transitions within the same ion, particularly $\rm [OIII](j_{\lambda 4959}+j_{\lambda 5007})/j_{\lambda 4363}$ and assuming a temperature of the gas \citep{Osterbrock:2006}.  However, while the measured [OIII] ratios are well-within the typical values for AGN NLRs, we note that $n_e$ is strongly dependent on the NLR temperature, and photoionization models imply NLRs may have multiple zones with different temperatures and densities \citep{Krolik:1999}.  Therefore, we have chosen to simply use the typical range of densities in our estimates.  Given the range of distances for each system, the possible ratios between the distance of the `blue' component to that of the `red' component range from 1:3 to 2.5:1.  In no case does the ratio of distances imply one source is significantly further than the other, providing no strong evidence for just one source being associated with an AGN.

\emph{Two AGNs:}  It is generally agreed that galaxy mergers, through tidal interactions, are effective at enhancing SMBH activity by funneling gas and dust toward the central region of the new distribution of mass \citep{Hernquist:1989,Springel:2005}.  Therefore, if CXOJ1426+35 did experience a wet merger, then the merger would have increased the probability for the two SMBHs to be active as gas would be forced toward the potential well of each SMBH.  On these grounds a dual AGN scenario is highly plausible and would naturally explain the two emission line systems.  A potential piece of evidence against this hypothesis is the NIRC2-AO image which reveals only one source at rest-frame $1\micron$ as opposed to the two nuclei expected in the dual AGN scenario.  However, there are several properties of CXOJ1426+35 which point to a high level of internal extinction capable of obscuring an AGN continuum at rest-frame $1\micron$.  In particular, the high X-ray HR implies there is at least one AGN suffering significant absorption, e.g. $n_H=6\times 10^{23}$ cm$^{-2}$ (Section \ref{Chandra}) and modeling of the SED (Figure \ref{sed_vfv}) reveals that an AGN with comparable absorption would be virtually invisible at rest-frame $1\micron$ (Section \ref{sed_model}).  This AGN would dominate in the X-ray and IRAC bands, while a second, less obscured AGN would be associated with the source seen at $1\micron$.   We note that another candidate dual AGN \citep{Comerford:2011} also shows only a single component in AO imaging, but which may host two heavily obscured (Compton thick) AGN as revealed from $Chandra$ data.  As discussed in Section \ref{intro}, several confirmed dual AGN \citep{Guainazzi2005,Bianchi2008,Piconcelli2010,Koss:2011} provide examples of this scenario,  showing only one visible nucleus, with the other heavily obscured and identifiable only through X-ray observations.  Given the prevalence of heavy obscuration in other candidate and confirmed dual AGN, the lack of detection of two cores would not be surprising if CXOJ1426+35 in fact hosted a dual AGN.

\subsection{Multiple Scenarios}

Finally, we note that more than one of these physical mechanisms may be present.  In particular, we note the possible coexistence of a galaxy merger and AGN outflows.  This follows from the morphology of CXOJ1426+35, which suggests a possible merger, and from the notion that galaxy mergers are potential instigators of enhanced SMBH activity.  If the merger has resulted in an AGN with a relatively high Eddington ratio (e.g., $>0.25$ and perhaps close to the Eddington limit), the accretion disk radiation pressure may be capable of producing the double-peaked narrow lines as discussed in Section \ref{outflows}.  While an ionization stratification is not apparent given the current data, we could not rule one out for either system.  This scenario is consistent with both the morphological suggestion of a recent merger while allowing for the existence of strong outflows driven by the radiation pressure of AGN accretion disks.

\subsection{Summary}

We have discussed six separate physical mechanisms that are each capable of producing double-peaked emission lines, and we have discussed the likelihood of each scenario based on the available data.  We find that a superposition, even within a galaxy cluster, is highly unlikely given the small physical and velocity separation, and that the rotating accretion disk scenario is highly unlikely given the several kiloparsec separation.  We find that a jet/cloud interaction is also highly unlikely simply because it is not a strong radio source.  The scenario of a NLR with an disk-shaped component can naturally explain the double-peaks, but the disagreement between the line ratios and line strengths of both systems is not expected in that scenario.  We find that the biconical outflow from an accretion disk can not be ruled out, and that the primary limitation is the power of the accretion disk.  A recent galactic merger would potentially make the biconical outflow a more likely explanation, although there is no clear line stratification.  In the case of a single AGN, the SED modeling implies there might also be a very young stellar population present which accounts for the UV excess.  A merger would also increase the possibility of a single AGN illuminating the NLR of each galaxy, though this scenario may not be likely given the short free-fall timescale and the similar distances of each cloud based on a single ionizing source.  A more likely scenario is that there are actually two AGN present.  The lack of two nuclei in the NIR AO image provides possible evidence against a second AGN, but this may be naturally explained by the high obscuration, inferred from both the X-ray data and the SED modeling, which would obscure the continuum radiation.  This scenario is also consistent with the several kiloparsec separation of the AGN components, since many of the dual AGNs already identified have comparable separations, and it is consistent with the AGN-like line ratios.       

\section{Tests for Dual and Binary SMBHs}
\label{tests}

Based upon limited data, determining the true nature of many candidate dual and binary SMBH systems has proven to be difficult given the attractiveness of alternative explanations.  Among the current handful of candidates there is a variety of data and interpretations, many of which have been tested and often with ambiguous results.  Here we use CXOJ1426+35, in comparison with other candidates, to review how these candidates can be tested.

\subsection{One-Dimensional Spectroscopy}

Given the availability of spectroscopic databases that include large samples of AGN and quasars, emission line diagnostics are commonly used to select the most promising double-peaked emission line candidates for follow-up observations.  From one-dimensional spectra alone (i.e. SDSS fiber spectra), much can be done to address the likelihood of a dual/binary AGN.  Emission line ratios can provide information about the source of the ionizing continuum radiation \citep{Baldwin1981,Kauffmann:2003}, and careful examination of spectra may also reveal asymmetries, velocity-offsets, or multiple components, thereby providing valuable information about the kinematics of the NLR gas.  We have performed these analyses for CXOJ1426+35 in the above discussion, and they have been done for all of the other candidate dual/binary SMBHs yet identified, providing valuable information and resulting in a variety of interpretations.  For example, in J0927+2943, close examination of the optical spectra (which features two prominent emission line systems, one with just narrow lines and the other with broad and narrow lines) has allowed for multiple models to be put forth, including a recoiling SMBH \citep{Komossa08a}, a binary SMBH \citep{Bogdanovic09b,Dotti09}, a superposition \citep{Shields2009} and a large and small galaxy interacting near the center of a rich cluster \citep{Heckman2009}.  As another example, in J1316+1753 \citet{XK09} identified two NLRs and at an intermediate redshift there is a broad component to each emission line, which the authors discuss may be evidence of a starburst-driven outflow that was triggered by a merger.  

We also note that spectroscopic observations over multiple epochs could potentially reveal any orbital motion in a close binary SMBH system.  Though CXOJ1426+35 does not fall into this class as the implied separation is much too large, candidate close binary SMBHs could be identified in this manner.  This test has been performed for both J0927+2943 \citep{Vivek2009} and J1536+0441 \citep{Chornock10} which were identified as candidate sub-parsec binaries with separations of $\sim0.34$ pc \citep{Dotti09} and $\sim0.1$ pc \citep{BL09}, respectively.  In both cases, the acceleration is found to be zero within the errors.  However, it has been pointed out that such candidates would be preferentially identified when their radial accelerations are smallest  \citep{Bogdanovic09a}.              

\subsection{Two-Dimensional Spectroscopy}

By providing information about the spatial separation of the two emission line components, two-dimensional spectroscopy has proven to be highly effective at identifying the most promising dual AGN candidates from the much larger sample of double-peaked narrow emission line sources.  In this way, CXOJ1426+35 has been shown to be a strong dual AGN candidate based on the two emission line components spatially separated and with a velocity difference of several hundred km s$^{-1}$.  At lower redshifts, there are many other similar candidates for which two-dimensional spectroscopy reveals two clearly separated components \citep[e.g.][]{Comerford2009a,Comerford2009b}.  Two-dimensional spectroscopy ruled out the close binary hypothesis for J0927+2943, but showed that it may still be a system of two merging galaxies \citep{Vivek2009}.  A large sample of double-peaked AGN with follow-up two-dimensional spectroscopy will increase the number of candidate dual AGN (Comerford et al., in prep). However, as made evident in our analysis of CXOJ1426+35, two-dimensional spectroscopy alone can not confirm a dual AGN.

\subsection{Morphology}

Morphology can provide powerful evidence for or against the dual AGN hypothesis.  A number of objects which show double-peaked emission lines have been imaged in optical and near-IR bands, where the results have shown morphological evidence of interactions/mergers such as tidal tails and multiple nuclei \citep{Junkkarinen2001,Comerford2009b,Green2010,Liu2010b}.  From $HST$ imaging, the galaxy COSMOS J100043.15+020637.2 has been established as hosting either a dual \citep{Comerford2009a} or a recoiling SMBH \citep{Civano2010}.  In the case of EGSD2 J142033.66+525917.5 \citep{Gerke2007}, an $HST$ ACS image shows morphological evidence of a possible interaction, but no clear evidence for multiple nuclei.  \citet{Rosario:2011} and \citet{Shen:2011} have obtained NIR imaging of several [OIII] double-peaked AGN where the results have shown a variety of possible origins of the double-peaked narrow emission lines, including some sources with morphological evidence suggesting the most likely interpretation is a dual AGN.  \citet{Fu2010} present images for a sample of several such candidates which reveal multiple nuclei in some cases, though they caution that spatially resolved spectroscopy is required to confirm any case as a true dual AGN.  In \citet{McGurk:2011} integral field spectroscopy was able to confirm a dual AGN.  That source was confirmed in the same way in \citet{Fu:2011b}, though their sample revealed that, out of 42 sources, in all but two the double-peaked lines were produced by gas kinematics.  These studies have highlighted the usefulness of integral field spectroscopy in eliminating the ambiguities left by double-peaked emission lines and even high-resolution imaging.  In the case of  CXOJ1426+35, NIR AO imaging revealed only a single nucleus at a rest-frame effective wavelength of $1\micron$, providing possible evidence against the dual AGN scenario.  As with other dual AGN candidates, integral field spectroscopy will be useful for CXOJ1426+35.  

In the case of heavy nuclear obscuration, however, detection of double point sources in X-ray images of close ULIRGS has been effective at identifying dual AGN, e.g., NGC 6240  \citep{Komossa2003}, ESO509-IG066 \citep{Guainazzi2005}, 3C 75 \citep{Hudson2006}, Mrk 436 \citep{Bianchi2008}, and IRAS 20210+1121 \citep{Piconcelli2010}.  The advantage of this method of detection is that hard X-rays are much less attenuated by intervening material in the galaxy, and recent mergers are known to have high column densities in the nuclear regions.  This might be the case for CXOJ1426+35, as dust may have a strong role in hiding an AGN near $1 \micron$, whereas two AGN would be revealed from hard X-rays.  Though CXOJ1426+35 is unresolved in our image, this is due to the off-axis detection and the resolution of the ACIS detector.  With the HRC, the two X-ray components could potentially be resolved in CXOJ1426+35. 

\section{Conclusions}
\label{conclusions}

We have analyzed all available data, new and archival, of the source CXOJ1426+35 which has an intriguing spectrum showing double-peaked optical/UV emission lines.  The two emission line components, separated spatially by 0.69" (5.5 kpc) and in velocity-space by 690 km s$^{-1}$, make CXOJ1426+35 similar to other candidate dual AGN, though it is unique in that it is the highest-redshift candidate dual AGN yet identified.  The optical/NIR imaging shows an elongated and ``lumpy'' morphological structure suggestive of a disturbance, whereas the AO image shows only a single source.  We have discussed several interpretations of the physical nature of CXOJ1426+35, including a chance superposition, a jet/cloud interaction, line emission from a rotating accretion disk, an unusual NLR geometry, a strong biconical outflow from an accretion disk, a dual SMBH and a combination of these scenarios.  We find that the only plausible scenarios are a strong outflow, a dual SMBH system, or a combination thereof.  If CXOJ1426+35 recently underwent a merger, an increased Eddington ratio may have resulted, making the disk outflow scenario a possible explanation.  Alternatively, a merger may have resulted in the actual case of a dual AGN, or, less likely, a single AGN illuminating the NLRs of both SMBHs.  Regardless, CXOJ1426+35 is  most likely a merger remnant.  

Finally, we compared CXOJ1426+35 to other candidates with the aim of reviewing methods of testing the dual AGN hypothesis.  Analysis of single-epoch 1D and 2D spectra provides crucial information about the ionizing continuum and the velocity and spatial separation of the two emission line components, but high-resolution imaging and multi-epoch spectra have proven most effective at confirming or rejecting the dual/binary SMBH hypothesis.  CXOJ1426+35 represents an interesting example of a dual AGN candidate, with many properties similar to other candidates in the literature, but also with several unique characteristics.  In particular, CXOJ1426+35 is the highest redshift dual AGN candidate to-date, at a redshift of $z=1.175$ when galaxy mergers were more frequent, making the dual AGN hypothesis especially appealing.  This makes it a fascinating addition and will help in developing a better understanding of how to look for the potential signatures of dual/binary SMBHs and of the often complex emission line regions of AGN.  

~\newline
The authors thank an anonymous referee for helpful comments that greatly improved the quality of the paper.  The authors gratefully acknowledge discussions on similar dual AGN candidates with close collaborators, especially Claud Lacy, Daniel Kennefick, Julia Kennefick and Joel Berrier.  RSB is also grateful to DS for funding an extended visit to JPL in Summer 2010.  AG acknowledges support from the National Science Foundation under grant AST-0708490.  This research was partially carried out at the Jet Propulsion Laboratory/California Institute of Technology, and was sponsored by the Strategic University Research Partnership Program and the National Aeronautics and Space Administration.  This work is based in part on observations obtained at the W. M.  Keck Observatory, which is operated as a scientific partnership among the California Institute of Technology, the University of California and the National Aeronautics and Space Administration.  The Observatory was made possible by the generous financial support of the W. M.  Keck Foundation.  This work is based in part on observations made with the \emph{Spitzer Space Telescope}, which is operated by the Jet Propulsion Laboratory, California Institute of Technology under a contract with NASA.  Support for this work was provided by NASA through an award issued by JPL/Caltech.  This work is based in part on observations obtained with the \emph{Chandra X-ray Observatory} ($CXO$), under contract SV4-74018, A31 with the Smithsonian Astrophysical Observatory which operates the $CXO$ for NASA.  This work makes use of image data from the NOAO Deep Wide-Field Survey (NDWFS) as distributed by the NOAO Science Archive.  NOAO is operated by the Association of Universities for Research in Astronomy (AURA), Inc., under a cooperative agreement with the National Science Foundation. This paper would not have been possible without the efforts of the $Spitzer$, $Chandra$, and Keck support staff.

\begin{deluxetable*}{lccccc}
\tabletypesize{\footnotesize}
\tablecolumns{6}
\tablecaption{Photometric data for CXOJ1426+35.}
\tablehead{
  \colhead{Obs. Bandpass} &
  \colhead{Rest Bandpass} &
  \colhead{$m_{\rm AB}$} &
  \colhead{$m_{\rm Vega}$} &
  \colhead{$F_{\nu}$ ($\mu$Jy)\tablenotemark{e}} &
  \colhead{Detector/Instrument}
}
\startdata \
2-10 keV & 13.0 keV & ... & ... & $2.81^{+0.33}_{-0.30} \times 10^{-3}$ & \emph{Chandra}/ACIS \\ \
0.5-2 keV & 2.7 keV & ... & ... & $4.44^{+4.01}_{-2.11} \times 10^{-5}$ & \emph{Chandra}/ACIS \\ \
2271 \AA\tablenotemark{a} & 1060 \AA & 22.24$\pm0.02$ & 20.57$\pm0.02$ & 4.63$^{0.10}_{0.09}$ & \emph{GALEX}/NUV \\ \
\emph{u}\tablenotemark{b} & 1630 \AA & 21.56$\pm0.23$ & 20.56$\pm0.23$ &  8.95$^{+2.01}_{-1.64}$ & SDSS \\ \
\emph{g}\tablenotemark{b} & 2190 \AA & 21.78$\pm0.09$ & 21.87$\pm0.09$ & 7.05$^{+0.54}_{-0.50}$ & SDSS \\ \
\emph{r}\tablenotemark{b} & 2865 \AA & 21.26$\pm0.09$ & 21.10$\pm0.09$ & 11.38$^{+0.87}_{-0.81}$ & SDSS \\ \
\emph{i}\tablenotemark{b} & 3505 \AA & 20.71$\pm0.09$ & 20.32$\pm0.09$ & 18.88$^{+1.44}_{-1.34}$ & SDSS \\ \
\emph{z}\tablenotemark{b} & 4200 \AA & 20.50$\pm0.26$ & 19.98$\pm0.26$ & 22.49$^{+6.09}_{-4.79}$ & SDSS \\ \
\emph{J}\tablenotemark{c} & 5975 \AA & 20.37$\pm0.09$ & 19.46$\pm0.09$ & 26.16$^{+2.72}_{-2.47}$ & NOAO/NEWFIRM \\ \
\emph{H}\tablenotemark{c} & 7585 \AA & 20.41$\pm0.28$ & 19.02$\pm0.28$ & 25.34$^{+8.13}_{-6.17}$ & NOAO/NEWFIRM \\ \
\emph{K$_s$}\tablenotemark{c} & 1.0$\micron$ & 20.10$\pm0.58$ & 18.25$\pm0.58$ & 33.53$^{+24.69}_{-14.23}$ & NOAO/NEWFIRM \\ \
3.6$\micron$\tablenotemark{d} & 1.7$\micron$ & 19.76$\pm0.02$ & 16.97$\pm0.02$ & 45.77$^{+1.53}_{-1.49}$ & \emph{Spitzer}/IRAC \\ \
4.5$\micron$\tablenotemark{d} & 2.1$\micron$ & 19.47$\pm0.02$ & 16.21$\pm0.02$ & 58.96$^{+1.97}_{-1.91}$ & \emph{Spitzer}/IRAC \\ \
5.8$\micron$\tablenotemark{d} & 2.7$\micron$ & 18.87$\pm0.06$ & 15.14$\pm0.06$ & 101.09$^{+7.32}_{-6.85}$ & \emph{Spitzer}/IRAC \\ \
8.0$\micron$\tablenotemark{d} & 3.7$\micron$ & 17.97$\pm0.03$ & 13.57$\pm0.03$ & 242.24$^{+10.24}_{-9.87}$ & \emph{Spitzer}/IRAC \\ \
\enddata
\tablenotetext{a}{Magnitude determined for the 3.8$^{\prime \prime}$ radius aperture corrected to total magnitude (see Section \ref{uv}).}
\tablenotetext{b}{SDSS Model magnitudes (see Section \ref{optical_nir}).}
\tablenotetext{c}{Magnitudes are for a Kron-like elliptical aperture (see Section \ref{optical_nir}).}
\tablenotetext{d}{Magnitudes determined for the 4$^{\prime \prime}$ radius aperture corrected to total magnitudes (see Section \ref{mid-ir}).}
\tablenotetext{e}{The \emph{Chandra} flux densities are not absorption-corrected. \\ }
\label{phot}
\end{deluxetable*} 

\begin{deluxetable*}{lccccccc}
\tabletypesize{\footnotesize}
\tablecaption{Fluxes, FWHMs, redshifts and velocity-splittings for emission lines in the both systems.}
\tablehead{
 \colhead{~} &
 \multicolumn{3}{c}{Blue System} &
 \multicolumn{3}{c}{Red System} \\
 \colhead{Line} &
 \colhead{Flux} &
 \colhead{FWHM} &
 \colhead{$z$} &
 \colhead{Flux} &
 \colhead{FWHM} &
 \colhead{$z$} &
 \colhead{Velocity-splits}
}
\startdata \
CIV1549 & 23.30$\pm6.23$ & 1120.95\tablenotemark{a} & 1.1718\tablenotemark{c} & 109.04$\pm9.85$ & 1120.95$\pm85.65$ & 1.1776 & 797.37$\pm75.49$ \\ \
HeII1640 & 20.70$\pm6.50$ & 818.90$\pm162.03$ & 1.1715 & 29.52$\pm6.58$ & 872.33$\pm141.95$ & 1.1779 & 882.47$\pm157.33$ \\ \
[OIII]1663 & 1.65$\pm0.47$ & 555.68\tablenotemark{b} & 1.1720\tablenotemark{b} & 3.10$\pm0.54$ & 474.17\tablenotemark{b} & 1.1770\tablenotemark{b} & 690.95\tablenotemark{b} \\ \ 
CIII]1909 & 14.72$\pm6.19$ & 925.58$\pm154.08$ & 1.1687 & 14.34$\pm6.22$ & 1003.31$\pm192.97$ & 1.1745 & 971.68$\pm236.55$ \\ \
CII]2326 & 3.30$\pm1.90$ & 880.83$\pm63.68$ & 1.1718 & 2.32$\pm1.98$ & 901.53$\pm239.35$ & 1.7758 & 799.66$\pm462.92$ \\ \
[NeV]3426 & 19.14$\pm1.25$ & 590.33$\pm39.64$ & 1.1715 & 17.48$\pm1.20$ & 509.30$\pm36$ & 1.1773 & 795.14$\pm28.90$ \\ \
[OII]3727 & 178.05$\pm76.04$ & 719.33$\pm170.34$ & 1.1722 & 80.10$\pm82.79$ & 697.81$\pm237.48$ & 1.1771 & 681.88$\pm333.59$ \\ \
[NeIII]3869 & 67.97$\pm5.73$ & 615.60$\pm63.09$ & 1.1718 & 37.06$\pm6.54$ & 542.78$\pm97.99$ & 1.1772 & 736.66$\pm49.73$ \\ \
[NeIII]3968 & 36.13$\pm2.27$ & 615.60\tablenotemark{c} & 1.1718\tablenotemark{c} & 19.35$\pm1.61$ & 542.78\tablenotemark{c} &  1.1772\tablenotemark{c} & 736.66\tablenotemark{c} \\ \
H$\gamma$ & 59.11$\pm8.21$ & 587.03\tablenotemark{d} & 1.1724\tablenotemark{d} & 33.46$\pm6.39$ & 474.17\tablenotemark{d} & 1.1766\tablenotemark{d} & 579.46\tablenotemark{d} \\ \
[OIII]4363 & 26.47$\pm5.04$ & 555.68\tablenotemark{b} & 1.1720\tablenotemark{b} & 12.34$\pm3.46$ & 474.17\tablenotemark{b} & 1.1770\tablenotemark{b} & 691.04\tablenotemark{b} \\ \
H$\beta$ & 178.67$\pm63.15$ & 587.03$\pm274.57$ & 1.1724 & $\leq 79.98$ & 474.17\tablenotemark{b} & 1.1766 & 579.46$\pm156.43$ \\ \
[OIII]4959 & 511.11$\pm108.30$ & 487.90$\pm59.76$ & 1.1721 & 186.61$\pm40.34$ & 474.17\tablenotemark{b} & 1.1770\tablenotemark{b} & 675.94$\pm44.75$ \\ \
[OIII]5007 & 1581.48$\pm132.82$ & 555.68$\pm19$ & 1.1720 & 477.93 & 474.17$\pm45.55$ & 1.1770 & 690.95$\pm44.82$ \\ \
\enddata
\tablecomments{Fluxes are in units of $10^{-17}$ erg s$^{-1}$ cm$^{-2}$ \AA$^{-1}$, and FWHMs and velocity-splttings relative to $z=1.175$ are in units of km s$^{-1}$.  All errors correspond to $1\sigma$ uncertainties.}
\tablenotetext{a}{Fixed to the `red' CIV1549 component.}
\tablenotetext{b}{Fixed to [OIII]5007.}
\tablenotetext{c}{Fixed to the value for [NeIII]3869.} 
\tablenotetext{d}{Fixed to the value for H$\beta$.  \\  }
\label{lines}
\end{deluxetable*}


\begin{thebibliography}{125}
\expandafter\ifx\csname natexlab\endcsname\relax\def\natexlab#1{#1}\fi

\bibitem[{{Abazajian} {et~al.}(2009){Abazajian}, {Adelman-McCarthy},
  {Ag{\"u}eros}, {Allam}, {Allende Prieto}, {An}, {Anderson}, {Anderson},
  {Annis}, {Bahcall}, {Bailer-Jones}, {Barentine}, {Bassett}, {Becker},
  {Beers}, {Bell}, {Belokurov}, {Berlind}, {Berman}, {Bernardi}, {Bickerton},
  {Bizyaev}, {Blakeslee}, {Blanton}, {Bochanski}, {Boroski}, {Brewington},
  {Brinchmann}, {Brinkmann}, {Brunner}, {Budav{\'a}ri}, {Carey}, {Carliles},
  {Carr}, {Castander}, {Cinabro}, {Connolly}, {Csabai}, {Cunha}, {Czarapata},
  {Davenport}, {de Haas}, {Dilday}, {Doi}, {Eisenstein}, {Evans}, {Evans},
  {Fan}, {Friedman}, {Frieman}, {Fukugita}, {G{\"a}nsicke}, {Gates},
  {Gillespie}, {Gilmore}, {Gonzalez}, {Gonzalez}, {Grebel}, {Gunn},
  {Gy{\"o}ry}, {Hall}, {Harding}, {Harris}, {Harvanek}, {Hawley}, {Hayes},
  {Heckman}, {Hendry}, {Hennessy}, {Hindsley}, {Hoblitt}, {Hogan}, {Hogg},
  {Holtzman}, {Hyde}, {Ichikawa}, {Ichikawa}, {Im}, {Ivezi{\'c}}, {Jester},
  {Jiang}, {Johnson}, {Jorgensen}, {Juri{\'c}}, {Kent}, {Kessler}, {Kleinman},
  {Knapp}, {Konishi}, {Kron}, {Krzesinski}, {Kuropatkin}, {Lampeitl},
  {Lebedeva}, {Lee}, {Lee}, {Leger}, {L{\'e}pine}, {Li}, {Lima}, {Lin}, {Long},
  {Loomis}, {Loveday}, {Lupton}, {Magnier}, {Malanushenko}, {Malanushenko},
  {Mandelbaum}, {Margon}, {Marriner}, {Mart{\'{\i}}nez-Delgado}, {Matsubara},
  {McGehee}, {McKay}, {Meiksin}, {Morrison}, {Mullally}, {Munn}, {Murphy},
  {Nash}, {Nebot}, {Neilsen}, {Newberg}, {Newman}, {Nichol}, {Nicinski},
  {Nieto-Santisteban}, {Nitta}, {Okamura}, {Oravetz}, {Ostriker}, {Owen},
  {Padmanabhan}, {Pan}, {Park}, {Pauls}, {Peoples}, {Percival}, {Pier}, {Pope},
  {Pourbaix}, {Price}, {Purger}, {Quinn}, {Raddick}, {Fiorentin}, {Richards},
  {Richmond}, {Riess}, {Rix}, {Rockosi}, {Sako}, {Schlegel}, {Schneider},
  {Scholz}, {Schreiber}, {Schwope}, {Seljak}, {Sesar}, {Sheldon}, {Shimasaku},
  {Sibley}, {Simmons}, {Sivarani}, {Smith}, {Smith}, {Smol{\v c}i{\'c}},
  {Snedden}, {Stebbins}, {Steinmetz}, {Stoughton}, {Strauss}, {Subba Rao},
  {Suto}, {Szalay}, {Szapudi}, {Szkody}, {Tanaka}, {Tegmark}, {Teodoro},
  {Thakar}, {Tremonti}, {Tucker}, {Uomoto}, {Vanden Berk}, {Vandenberg},
  {Vidrih}, {Vogeley}, {Voges}, {Vogt}, {Wadadekar}, {Watters}, {Weinberg},
  {West}, {White}, {Wilhite}, {Wonders}, {Yanny}, {Yocum}, {York}, {Zehavi},
  {Zibetti}, \& {Zucker}}]{Abazajian09}
{Abazajian}, K.~N., {et~al.} 2009, ApJ, 182, 543

\bibitem[{{Ashby} {et~al.}(2009){Ashby}, {Stern}, {Brodwin}, {Griffith},
  {Eisenhardt}, {Koz{\l}owski}, {Kochanek}, {Bock}, {Borys}, {Brand}, {Brown},
  {Cool}, {Cooray}, {Croft}, {Dey}, {Eisenstein}, {Gonzalez}, {Gorjian},
  {Grogin}, {Ivison}, {Jacob}, {Jannuzi}, {Mainzer}, {Moustakas},
  {R{\"o}ttgering}, {Seymour}, {Smith}, {Stanford}, {Stauffer}, {Sullivan},
  {van Breugel}, {Willner}, \& {Wright}}]{ashby2009}
{Ashby}, M.~L.~N., {et~al.} 2009, \apj, 701, 428

\bibitem[{{Assef} {et~al.}(2010){Assef}, {Kochanek}, {Brodwin}, {Cool},
  {Forman}, {Gonzalez}, {Hickox}, {Jones}, {Le Floc'h}, {Moustakas}, {Murray},
  \& {Stern}}]{assef2010}
{Assef}, R.~J., {et~al.} 2010, AJ, 713, 970

\bibitem[{{Baldwin} {et~al.}(1981){Baldwin}, {Phillips}, \&
  {Terlevich}}]{Baldwin1981}
{Baldwin}, J.~A., {Phillips}, M.~M., \& {Terlevich}, R. 1981, \pasp, 93, 5

\bibitem[{{Barrows} {et~al.}(2011){Barrows}, {Lacy}, {Kennefick}, {Kennefick},
  \& {Seigar}}]{Barrows:2011}
{Barrows}, R.~S., {Lacy}, C.~H.~S., {Kennefick}, D., {Kennefick}, J., \&
  {Seigar}, M.~S. 2011, \na, 16, 122

\bibitem[{{Begelman} {et~al.}(1980){Begelman}, {Blandford}, \&
  {Rees}}]{Begelman1980}
{Begelman}, M.~C., {Blandford}, R.~D., \& {Rees}, M.~J. 1980, \nat, 287, 307

\bibitem[{{Bell} {et~al.}(2003){Bell}, {McIntosh}, {Katz}, \&
  {Weinberg}}]{Bell:2003}
{Bell}, E.~F., {McIntosh}, D.~H., {Katz}, N., \& {Weinberg}, M.~D. 2003, \apjs,
  149, 289

\bibitem[{{Bennert} {et~al.}(2002){Bennert}, {Falcke}, {Schulz}, {Wilson}, \&
  {Wills}}]{Bennert:2002}
{Bennert}, N., {Falcke}, H., {Schulz}, H., {Wilson}, A.~S., \& {Wills}, B.~J.
  2002, \apjl, 574, L105

\bibitem[{{Bianchi} {et~al.}(2008){Bianchi}, {Chiaberge}, {Piconcelli},
  {Guainazzi}, \& {Matt}}]{Bianchi2008}
{Bianchi}, S., {Chiaberge}, M., {Piconcelli}, E., {Guainazzi}, M., \& {Matt},
  G. 2008, \mnras, 386, 105

\bibitem[{{Blanton} {et~al.}(2005){Blanton}, {Schlegel}, {Strauss},
  {Brinkmann}, {Finkbeiner}, {Fukugita}, {Gunn}, {Hogg}, {Ivezi{\'c}}, {Knapp},
  {Lupton}, {Munn}, {Schneider}, {Tegmark}, \& {Zehavi}}]{Blanton2005}
{Blanton}, M.~R., {et~al.} 2005, \aj, 129, 2562

\bibitem[{{Blustin} \& {Fabian}(2009)}]{Blustin:2009}
{Blustin}, A.~J., \& {Fabian}, A.~C. 2009, \mnras, 396, 1732

\bibitem[{{Bogdanovi{\'c}} {et~al.}(2009{\natexlab{a}}){Bogdanovi{\'c}},
  {Eracleous}, \& {Sigurdsson}}]{Bogdanovic09a}
{Bogdanovi{\'c}}, T., {Eracleous}, M., \& {Sigurdsson}, S. 2009{\natexlab{a}},
  NewA, 53, 113

\bibitem[{{Bogdanovi{\'c}} {et~al.}(2009{\natexlab{b}}){Bogdanovi{\'c}},
  {Eracleous}, \& {Sigurdsson}}]{Bogdanovic09b}
---. 2009{\natexlab{b}}, ApJ, 697, 288

\bibitem[{{Bohlin} {et~al.}(2001){Bohlin}, {Dickinson}, \&
  {Calzetti}}]{Bohlin2001}
{Bohlin}, R.~C., {Dickinson}, M.~E., \& {Calzetti}, D. 2001, \aj, 122, 2118

\bibitem[{{Boroson} \& {Lauer}(2009)}]{BL09}
{Boroson}, T.~A., \& {Lauer}, T.~R. 2009, Nature, 458, 53

\bibitem[{{Bridge} {et~al.}(2010){Bridge}, {Carlberg}, \&
  {Sullivan}}]{Bridge:2010}
{Bridge}, C.~R., {Carlberg}, R.~G., \& {Sullivan}, M. 2010, \apj, 709, 1067

\bibitem[{{Cardelli} {et~al.}(1989){Cardelli}, {Clayton}, \&
  {Mathis}}]{Cardelli:1989}
{Cardelli}, J.~A., {Clayton}, G.~C., \& {Mathis}, J.~S. 1989, \apj, 345, 245

\bibitem[{{Chen} \& {Halpern}(1989)}]{CH89}
{Chen}, K., \& {Halpern}, J.~P. 1989, ApJ, 344, 115

\bibitem[{{Chornock} {et~al.}(2010){Chornock}, {Bloom}, {Cenko}, {Filippenko},
  {Silverman}, {Hicks}, {Lawrence}, {Mendez}, {Rafelski}, \&
  {Wolfe}}]{Chornock10}
{Chornock}, R., {et~al.} 2010, ApJ, 709, L39

\bibitem[{{Civano} {et~al.}(2010){Civano}, {Elvis}, {Lanzuisi}, {Jahnke},
  {Zamorani}, {Blecha}, {Bongiorno}, {Brusa}, {Comastri}, {Hao}, {Leauthaud},
  {Loeb}, {Mainieri}, {Piconcelli}, {Salvato}, {Scoville}, {Trump}, {Vignali},
  {Aldcroft}, {Bolzonella}, {Bressert}, {Finoguenov}, {Fruscione}, {Koekemoer},
  {Cappelluti}, {Fiore}, {Giodini}, {Gilli}, {Impey}, {Lilly}, {Lusso},
  {Puccetti}, {Silverman}, {Aussel}, {Capak}, {Frayer}, {Le Floch},
  {McCracken}, {Sanders}, {Schiminovich}, \& {Taniguchi}}]{Civano2010}
{Civano}, F., {et~al.} 2010, \apj, 717, 209

\bibitem[{{Comerford} {et~al.}(2009{\natexlab{a}}){Comerford}, {Griffith},
  {Gerke}, {Cooper}, {Newman}, {Davis}, \& {Stern}}]{Comerford2009a}
{Comerford}, J.~M., {Griffith}, R.~L., {Gerke}, B.~F., {Cooper}, M.~C.,
  {Newman}, J.~A., {Davis}, M., \& {Stern}, D. 2009{\natexlab{a}}, \apjl, 702,
  L82

\bibitem[{{Comerford} {et~al.}(2011){Comerford}, {Pooley}, {Gerke}, \&
  {Madejski}}]{Comerford:2011}
{Comerford}, J.~M., {Pooley}, D., {Gerke}, B.~F., \& {Madejski}, G.~M. 2011,
  \apjl, 737, L19+

\bibitem[{{Comerford} {et~al.}(2009{\natexlab{b}}){Comerford}, {Gerke},
  {Newman}, {Davis}, {Yan}, {Cooper}, {Faber}, {Koo}, {Coil}, {Rosario}, \&
  {Dutton}}]{Comerford2009b}
{Comerford}, J.~M., {et~al.} 2009{\natexlab{b}}, \apj, 698, 956

\bibitem[{{Cooper} {et~al.}(2008){Cooper}, {Tremonti}, {Newman}, \&
  {Zabludoff}}]{Cooper:2008}
{Cooper}, M.~C., {Tremonti}, C.~A., {Newman}, J.~A., \& {Zabludoff}, A.~I.
  2008, \mnras, 390, 245

\bibitem[{{Daly}(1992)}]{Daly:1992}
{Daly}, R.~A. 1992, \apj, 399, 426

\bibitem[{{de Vries} {et~al.}(2002){de Vries}, {Morganti}, {R{\"o}ttgering},
  {Vermeulen}, {van Breugel}, {Rengelink}, \& {Jarvis}}]{deVries02}
{de Vries}, W.~H., {Morganti}, R., {R{\"o}ttgering}, H.~J.~A., {Vermeulen}, R.,
  {van Breugel}, W., {Rengelink}, R., \& {Jarvis}, M.~J. 2002, AJ, 123, 1784

\bibitem[{{De Young}(1981)}]{DeYoung:1981}
{De Young}, D.~S. 1981, \nat, 293, 43

\bibitem[{{Decarli} {et~al.}(2010){Decarli}, {Dotti}, {Montuori}, {Liimets}, \&
  {Ederoclite}}]{Decarli2010}
{Decarli}, R., {Dotti}, M., {Montuori}, C., {Liimets}, T., \& {Ederoclite}, A.
  2010, \apjl, 720, L93

\bibitem[{{Dotti} {et~al.}(2009){Dotti}, {Montuori}, {Decarli}, {Volonteri},
  {Colpi}, \& {Haardt}}]{Dotti09}
{Dotti}, M., {Montuori}, C., {Decarli}, R., {Volonteri}, M., {Colpi}, M., \&
  {Haardt}, F. 2009, MNRA, 398, L73

\bibitem[{{Dotti} \& {Ruszkowski}(2010)}]{Dotti:2010}
{Dotti}, M., \& {Ruszkowski}, M. 2010, \apjl, 713, L37

\bibitem[{{Dressler} {et~al.}(1997){Dressler}, {Oemler}, {Couch}, {Smail},
  {Ellis}, {Barger}, {Butcher}, {Poggianti}, \& {Sharples}}]{Dressler:1997}
{Dressler}, A., {et~al.} 1997, \apj, 490, 577

\bibitem[{{Everett}(2007)}]{Everett2007b}
{Everett}, J.~E. 2007, \apss, 311, 269

\bibitem[{{Everett} \& {Murray}(2007)}]{Everett2007a}
{Everett}, J.~E., \& {Murray}, N. 2007, \apj, 656, 93

\bibitem[{{Fazio} {et~al.}(2004){Fazio}, {Hora}, {Allen}, {Ashby}, {Barmby},
  {Deutsch}, {Huang}, {Kleiner}, {Marengo}, {Megeath}, {Melnick}, {Pahre},
  {Patten}, {Polizotti}, {Smith}, {Taylor}, {Wang}, {Willner}, {Hoffmann},
  {Pipher}, {Forrest}, {McMurty}, {McCreight}, {McKelvey}, {McMurray}, {Koch},
  {Moseley}, {Arendt}, {Mentzell}, {Marx}, {Losch}, {Mayman}, {Eichhorn},
  {Krebs}, {Jhabvala}, {Gezari}, {Fixsen}, {Flores}, {Shakoorzadeh}, {Jungo},
  {Hakun}, {Workman}, {Karpati}, {Kichak}, {Whitley}, {Mann}, {Tollestrup},
  {Eisenhardt}, {Stern}, {Gorjian}, {Bhattacharya}, {Carey}, {Nelson},
  {Glaccum}, {Lacy}, {Lowrance}, {Laine}, {Reach}, {Stauffer}, {Surace},
  {Wilson}, {Wright}, {Hoffman}, {Domingo}, \& {Cohen}}]{Fazio2004}
{Fazio}, G.~G., {et~al.} 2004, \apjs, 154, 10

\bibitem[{{Ferland}(1981)}]{Ferland:1981}
{Ferland}, G.~J. 1981, \apj, 249, 17

\bibitem[{{Fischer} {et~al.}(2011){Fischer}, {Crenshaw}, {Kraemer}, {Schmitt},
  {Mushotsky}, \& {Dunn}}]{Fischer:2011}
{Fischer}, T.~C., {Crenshaw}, D.~M., {Kraemer}, S.~B., {Schmitt}, H.~R.,
  {Mushotsky}, R.~F., \& {Dunn}, J.~P. 2011, \apj, 727, 71

\bibitem[{{Foreman} {et~al.}(2009){Foreman}, {Volonteri}, \&
  {Dotti}}]{Foreman2009}
{Foreman}, G., {Volonteri}, M., \& {Dotti}, M. 2009, \apj, 693, 1554

\bibitem[{{Fu} {et~al.}(2010){Fu}, {Myers}, {Djorgovski}, \& {Yan}}]{Fu2010}
{Fu}, H., {Myers}, A.~D., {Djorgovski}, S.~G., \& {Yan}, L. 2010, ArXiv
  e-prints

\bibitem[{{Fu} \& {Stockton}(2009)}]{Fu2009}
{Fu}, H., \& {Stockton}, A. 2009, \apj, 690, 953

\bibitem[{{Fu} {et~al.}(2011){Fu}, {Yan}, {Myers}, {Stockton}, {Djorgovski},
  {Aldering}, \& {Rich}}]{Fu:2011b}
{Fu}, H., {Yan}, L., {Myers}, A.~D., {Stockton}, A., {Djorgovski}, S.~G.,
  {Aldering}, G., \& {Rich}, J.~A. 2011, ArXiv e-prints

\bibitem[{{Garmire} {et~al.}(2003){Garmire}, {Bautz}, {Ford}, {Nousek}, \&
  {Ricker}}]{Garmire2003}
{Garmire}, G.~P., {Bautz}, M.~W., {Ford}, P.~G., {Nousek}, J.~A., \& {Ricker},
  Jr., G.~R. 2003, in Society of Photo-Optical Instrumentation Engineers (SPIE)
  Conference Series, Vol. 4851, Society of Photo-Optical Instrumentation
  Engineers (SPIE) Conference Series, ed. {J.~E.~Truemper \& H.~D.~Tananbaum},
  28--44

\bibitem[{{Gerke} {et~al.}(2007){Gerke}, {Newman}, {Lotz}, {Yan}, {Barmby},
  {Coil}, {Conselice}, {Ivison}, {Lin}, {Koo}, {Nandra}, {Salim}, {Small},
  {Weiner}, {Cooper}, {Davis}, {Faber}, \& {Guhathakurta}}]{Gerke2007}
{Gerke}, B.~F., {et~al.} 2007, \apjl, 660, L23

\bibitem[{{Green} {et~al.}(2010){Green}, {Myers}, {Barkhouse}, {Mulchaey},
  {Bennert}, {Cox}, \& {Aldcroft}}]{Green2010}
{Green}, P.~J., {Myers}, A.~D., {Barkhouse}, W.~A., {Mulchaey}, J.~S.,
  {Bennert}, V.~N., {Cox}, T.~J., \& {Aldcroft}, T.~L. 2010, \apj, 710, 1578

\bibitem[{{Guainazzi} {et~al.}(2005){Guainazzi}, {Piconcelli},
  {Jim{\'e}nez-Bail{\'o}n}, \& {Matt}}]{Guainazzi2005}
{Guainazzi}, M., {Piconcelli}, E., {Jim{\'e}nez-Bail{\'o}n}, E., \& {Matt}, G.
  2005, \aap, 429, L9

\bibitem[{{H{\"a}ring} \& {Rix}(2004)}]{Haring:2004}
{H{\"a}ring}, N., \& {Rix}, H. 2004, \apjl, 604, L89

\bibitem[{{Heckman} {et~al.}(2009){Heckman}, {Krolik}, {Moran}, {Schnittman},
  \& {Gezari}}]{Heckman2009}
{Heckman}, T.~M., {Krolik}, J.~H., {Moran}, S.~M., {Schnittman}, J., \&
  {Gezari}, S. 2009, \apj, 695, 363

\bibitem[{{Heckman} {et~al.}(1981){Heckman}, {Miley}, {van Breugel}, \&
  {Butcher}}]{Heckman1981}
{Heckman}, T.~M., {Miley}, G.~K., {van Breugel}, W.~J.~M., \& {Butcher}, H.~R.
  1981, \apj, 247, 403

\bibitem[{{Hennawi} {et~al.}(2006){Hennawi}, {Strauss}, {Oguri}, {Inada},
  {Richards}, {Pindor}, {Schneider}, {Becker}, {Gregg}, {Hall}, {Johnston},
  {Fan}, {Burles}, {Schlegel}, {Gunn}, {Lupton}, {Bahcall}, {Brunner}, \&
  {Brinkmann}}]{Hennawi2006}
{Hennawi}, J.~F., {et~al.} 2006, \aj, 131, 1

\bibitem[{{Hennawi} {et~al.}(2010){Hennawi}, {Myers}, {Shen}, {Strauss},
  {Djorgovski}, {Fan}, {Glikman}, {Mahabal}, {Martin}, {Richards}, {Schneider},
  \& {Shankar}}]{Hennawi2010}
---. 2010, \apj, 719, 1672

\bibitem[{{Hernquist}(1989)}]{Hernquist:1989}
{Hernquist}, L. 1989, \nat, 340, 687

\bibitem[{{Hickox} {et~al.}(2007){Hickox}, {Jones}, {Forman}, {Murray},
  {Brodwin}, {Brown}, {Eisenhardt}, {Stern}, {Kochanek}, {Eisenstein}, {Cool},
  {Jannuzi}, {Dey}, {Brand}, {Gorjian}, \& {Caldwell}}]{Hickox:2007}
{Hickox}, R.~C., {et~al.} 2007, \apj, 671, 1365

\bibitem[{{Holt} {et~al.}(2003){Holt}, {Tadhunter}, \& {Morganti}}]{Holt2003}
{Holt}, J., {Tadhunter}, C.~N., \& {Morganti}, R. 2003, \mnras, 342, 227

\bibitem[{{Holt} {et~al.}(2008){Holt}, {Tadhunter}, \& {Morganti}}]{Holt2008}
---. 2008, \mnras, 387, 639

\bibitem[{{Hopkins} {et~al.}(2005){Hopkins}, {Hernquist}, {Cox}, {Di Matteo},
  {Martini}, {Robertson}, \& {Springel}}]{Hopkins05}
{Hopkins}, P.~F., {Hernquist}, L., {Cox}, T.~J., {Di Matteo}, T., {Martini},
  P., {Robertson}, B., \& {Springel}, V. 2005, ApJ, 630, 705

\bibitem[{{Hopkins} {et~al.}(2008){Hopkins}, {Hernquist}, {Cox}, \& {Kere{\v
  s}}}]{Hopkins2008}
{Hopkins}, P.~F., {Hernquist}, L., {Cox}, T.~J., \& {Kere{\v s}}, D. 2008,
  \apjs, 175, 356

\bibitem[{{Hopkins} {et~al.}(2009){Hopkins}, {Hickox}, {Quataert}, \&
  {Hernquist}}]{Hopkins:2009}
{Hopkins}, P.~F., {Hickox}, R., {Quataert}, E., \& {Hernquist}, L. 2009,
  \mnras, 398, 333

\bibitem[{{Hopkins} {et~al.}(2007){Hopkins}, {Richards}, \&
  {Hernquist}}]{Hopkins2007}
{Hopkins}, P.~F., {Richards}, G.~T., \& {Hernquist}, L. 2007, \apj, 654, 731

\bibitem[{{Hudson} {et~al.}(2006){Hudson}, {Reiprich}, {Clarke}, \&
  {Sarazin}}]{Hudson2006}
{Hudson}, D.~S., {Reiprich}, T.~H., {Clarke}, T.~E., \& {Sarazin}, C.~L. 2006,
  \aap, 453, 433

\bibitem[{{Jannuzi} {et~al.}(2000){Jannuzi}, {Dey}, {Tiede}, {Brown}, \& {NDWFS
  Team}}]{Jannuzi2000}
{Jannuzi}, B.~T., {Dey}, A., {Tiede}, G.~P., {Brown}, M.~J.~I., \& {NDWFS
  Team}. 2000, in Bulletin of the American Astronomical Society, Vol.~32,
  Bulletin of the American Astronomical Society, 1528--+

\bibitem[{{Junkkarinen} {et~al.}(2001){Junkkarinen}, {Shields}, {Beaver},
  {Burbidge}, {Cohen}, {Hamann}, \& {Lyons}}]{Junkkarinen2001}
{Junkkarinen}, V., {Shields}, G.~A., {Beaver}, E.~A., {Burbidge}, E.~M.,
  {Cohen}, R.~D., {Hamann}, F., \& {Lyons}, R.~W. 2001, \apjl, 549, L155

\bibitem[{{Kauffmann} {et~al.}(2003){Kauffmann}, {Heckman}, {Tremonti},
  {Brinchmann}, {Charlot}, {White}, {Ridgway}, {Brinkmann}, {Fukugita}, {Hall},
  {Ivezi{\'c}}, {Richards}, \& {Schneider}}]{Kauffmann:2003}
{Kauffmann}, G., {et~al.} 2003, \mnras, 346, 1055

\bibitem[{{Kelly} {et~al.}(2010){Kelly}, {Vestergaard}, {Fan}, {Hopkins},
  {Hernquist}, \& {Siemiginowska}}]{Kelly:2010}
{Kelly}, B.~C., {Vestergaard}, M., {Fan}, X., {Hopkins}, P., {Hernquist}, L.,
  \& {Siemiginowska}, A. 2010, \apj, 719, 1315

\bibitem[{{Kewley} {et~al.}(2001){Kewley}, {Dopita}, {Sutherland}, {Heisler},
  \& {Trevena}}]{Kewley:2001}
{Kewley}, L.~J., {Dopita}, M.~A., {Sutherland}, R.~S., {Heisler}, C.~A., \&
  {Trevena}, J. 2001, \apj, 556, 121

\bibitem[{{Kewley} {et~al.}(2006){Kewley}, {Groves}, {Kauffmann}, \&
  {Heckman}}]{Kewley:2006}
{Kewley}, L.~J., {Groves}, B., {Kauffmann}, G., \& {Heckman}, T. 2006, \mnras,
  372, 961

\bibitem[{{King}(2010)}]{King:2010}
{King}, A.~R. 2010, \mnras, 408, L95

\bibitem[{{Kollmeier} {et~al.}(2006){Kollmeier}, {Onken}, {Kochanek}, {Gould},
  {Weinberg}, {Dietrich}, {Cool}, {Dey}, {Eisenstein}, {Jannuzi}, {Le Floc'h},
  \& {Stern}}]{Kollmeier2006}
{Kollmeier}, J.~A., {et~al.} 2006, \apj, 648, 128

\bibitem[{{Komossa} {et~al.}(2003){Komossa}, {Burwitz}, {Hasinger}, {Predehl},
  {Kaastra}, \& {Ikebe}}]{Komossa2003}
{Komossa}, S., {Burwitz}, V., {Hasinger}, G., {Predehl}, P., {Kaastra}, J.~S.,
  \& {Ikebe}, Y. 2003, \apjl, 582, L15

\bibitem[{{Komossa} {et~al.}(2006){Komossa}, {Voges}, {Adorf}, {Xu}, {Mathur},
  \& {Anderson}}]{Komossa:2006}
{Komossa}, S., {Voges}, W., {Adorf}, H., {Xu}, D., {Mathur}, S., \& {Anderson},
  S.~F. 2006, \apj, 639, 710

\bibitem[{{Komossa} {et~al.}(2008{\natexlab{a}}){Komossa}, {Xu}, {Zhou},
  {Storchi-Bergmann}, \& {Binette}}]{Komossa2008b}
{Komossa}, S., {Xu}, D., {Zhou}, H., {Storchi-Bergmann}, T., \& {Binette}, L.
  2008{\natexlab{a}}, \apj, 680, 926

\bibitem[{{Komossa} {et~al.}(2008{\natexlab{b}}){Komossa}, {Zhou}, \&
  {Lu}}]{Komossa08a}
{Komossa}, S., {Zhou}, H., \& {Lu}, H. 2008{\natexlab{b}}, ApJ, 678, L81

\bibitem[{{Koss} {et~al.}(2011){Koss}, {Mushotzky}, {Treister}, {Veilleux},
  {Vasudevan}, {Miller}, {Sanders}, {Schawinski}, \& {Trippe}}]{Koss:2011}
{Koss}, M., {et~al.} 2011, \apjl, 735, L42+

\bibitem[{{Koz{\l}owski} {et~al.}(2010){Koz{\l}owski}, {Kochanek}, {Stern},
  {Prieto}, \& {Stanek}}]{Kozlowski2010}
{Koz{\l}owski}, S., {Kochanek}, C.~S., {Stern}, D., {Prieto}, J.~L., \&
  {Stanek}, K.~Z. 2010, Central Bureau Electronic Telegrams, 2392, 1

\bibitem[{{Kriss}(1994)}]{Kriss94}
{Kriss}, G. 1994, Astronomical Data Analysis Software and Systems, 3, 437

\bibitem[{{Krolik}(1999)}]{Krolik:1999}
{Krolik}, J.~H. 1999, {Active galactic nuclei : from the central black hole to
  the galactic environment}, ed. {Krolik, J.~H.}

\bibitem[{{Kurucz}(1993)}]{Kurucz1993}
{Kurucz}, R.~L. 1993, VizieR Online Data Catalog, 6039, 0

\bibitem[{{Lamastra} {et~al.}(2009){Lamastra}, {Bianchi}, {Matt}, {Perola},
  {Barcons}, \& {Carrera}}]{Lamastra:2009}
{Lamastra}, A., {Bianchi}, S., {Matt}, G., {Perola}, G.~C., {Barcons}, X., \&
  {Carrera}, F.~J. 2009, \aap, 504, 73

\bibitem[{{Liu} {et~al.}(2010{\natexlab{a}}){Liu}, {Greene}, {Shen}, \&
  {Strauss}}]{Liu2010b}
{Liu}, X., {Greene}, J.~E., {Shen}, Y., \& {Strauss}, M.~A. 2010{\natexlab{a}},
  \apjl, 715, L30

\bibitem[{{Liu} {et~al.}(2010{\natexlab{b}}){Liu}, {Shen}, {Strauss}, \&
  {Greene}}]{Liu2010a}
{Liu}, X., {Shen}, Y., {Strauss}, M.~A., \& {Greene}, J.~E. 2010{\natexlab{b}},
  \apj, 708, 427

\bibitem[{{Madau} {et~al.}(1998){Madau}, {Pozzetti}, \&
  {Dickinson}}]{Madau:1998}
{Madau}, P., {Pozzetti}, L., \& {Dickinson}, M. 1998, \apj, 498, 106

\bibitem[{{Maiolino} {et~al.}(2001){Maiolino}, {Marconi}, {Salvati},
  {Risaliti}, {Severgnini}, {Oliva}, {La Franca}, \& {Vanzi}}]{Maiolino:2001}
{Maiolino}, R., {Marconi}, A., {Salvati}, M., {Risaliti}, G., {Severgnini}, P.,
  {Oliva}, E., {La Franca}, F., \& {Vanzi}, L. 2001, \aap, 365, 28

\bibitem[{{Martin} {et~al.}(2003){Martin}, {Barlow}, {Barnhart}, {Bianchi},
  {Blakkolb}, {Bruno}, {Bushman}, {Byun}, {Chiville}, {Conrow}, {Cooke},
  {Donas}, {Fanson}, {Forster}, {Friedman}, {Grange}, {Griffiths}, {Heckman},
  {Lee}, {Jelinsky}, {Kim}, {Lee}, {Lee}, {Liu}, {Madore}, {Malina}, {Mazer},
  {McLean}, {Milliard}, {Mitchell}, {Morais}, {Morrissey}, {Neff}, {Raison},
  {Randall}, {Rich}, {Schiminovich}, {Schmitigal}, {Sen}, {Siegmund}, {Small},
  {Stock}, {Surber}, {Szalay}, {Vaughan}, {Weigand}, {Welsh}, {Wu}, {Wyder},
  {Xu}, \& {Zsoldas}}]{Martin2003}
{Martin}, C., {et~al.} 2003, in Society of Photo-Optical Instrumentation
  Engineers (SPIE) Conference Series, Vol. 4854, Society of Photo-Optical
  Instrumentation Engineers (SPIE) Conference Series, ed. {J.~C.~Blades \&
  O.~H.~W.~Siegmund}, 336--350

\bibitem[{{Maxfield} {et~al.}(2002){Maxfield}, {Spinrad}, {Stern}, {Dey}, \&
  {Dickinson}}]{Maxfield2002}
{Maxfield}, L., {Spinrad}, H., {Stern}, D., {Dey}, A., \& {Dickinson}, M. 2002,
  \aj, 123, 2321

\bibitem[{{McCarthy}(1993)}]{McCarthy1993}
{McCarthy}, P.~J. 1993, \araa, 31, 639

\bibitem[{{McGurk} {et~al.}(2011){McGurk}, {Max}, {Rosario}, {Shields},
  {Smith}, \& {Wright}}]{McGurk:2011}
{McGurk}, R.~C., {Max}, C.~E., {Rosario}, D.~J., {Shields}, G.~A., {Smith},
  K.~L., \& {Wright}, S.~A. 2011, ArXiv e-prints

\bibitem[{{McLean} {et~al.}(1998){McLean}, {Becklin}, {Bendiksen}, {Brims},
  {Canfield}, {Figer}, {Graham}, {Hare}, {Lacayanga}, {Larkin}, {Larson},
  {Levenson}, {Magnone}, {Teplitz}, \& {Wong}}]{McLean1998}
{McLean}, I.~S., {et~al.} 1998, in Society of Photo-Optical Instrumentation
  Engineers (SPIE) Conference Series, Vol. 3354, Society of Photo-Optical
  Instrumentation Engineers (SPIE) Conference Series, ed. {A.~M.~Fowler},
  566--578

\bibitem[{{Miley} \& {De Breuck}(2008)}]{Miley:DeBreuck:2008}
{Miley}, G., \& {De Breuck}, C. 2008, \aapr, 15, 67

\bibitem[{{Milosavljevi{\'c}} \& {Merritt}(2003)}]{Milosavljevic2003}
{Milosavljevi{\'c}}, M., \& {Merritt}, D. 2003, \apj, 596, 860

\bibitem[{{Morrissey} {et~al.}(2007){Morrissey}, {Conrow}, {Barlow}, {Small},
  {Seibert}, {Wyder}, {Budav{\'a}ri}, {Arnouts}, {Friedman}, {Forster},
  {Martin}, {Neff}, {Schiminovich}, {Bianchi}, {Donas}, {Heckman}, {Lee},
  {Madore}, {Milliard}, {Rich}, {Szalay}, {Welsh}, \& {Yi}}]{Morrissey2007}
{Morrissey}, P., {et~al.} 2007, \apjs, 173, 682

\bibitem[{{Myers} {et~al.}(2007){Myers}, {Brunner}, {Richards}, {Nichol},
  {Schneider}, \& {Bahcall}}]{Myers2007}
{Myers}, A.~D., {Brunner}, R.~J., {Richards}, G.~T., {Nichol}, R.~C.,
  {Schneider}, D.~P., \& {Bahcall}, N.~A. 2007, \apj, 658, 99

\bibitem[{{Myers} {et~al.}(2008){Myers}, {Richards}, {Brunner}, {Schneider},
  {Strand}, {Hall}, {Blomquist}, \& {York}}]{Myers2008}
{Myers}, A.~D., {Richards}, G.~T., {Brunner}, R.~J., {Schneider}, D.~P.,
  {Strand}, N.~E., {Hall}, P.~B., {Blomquist}, J.~A., \& {York}, D.~G. 2008,
  \apj, 678, 635

\bibitem[{{Nesvadba} {et~al.}(2006){Nesvadba}, {Lehnert}, {Eisenhauer},
  {Gilbert}, {Tecza}, \& {Abuter}}]{Nesvadba:2006}
{Nesvadba}, N.~P.~H., {Lehnert}, M.~D., {Eisenhauer}, F., {Gilbert}, A.,
  {Tecza}, M., \& {Abuter}, R. 2006, \apj, 650, 693

\bibitem[{{Oke} {et~al.}(1995){Oke}, {Cohen}, {Carr}, {Cromer}, {Dingizian},
  {Harris}, {Labrecque}, {Lucinio}, {Schaal}, {Epps}, \& {Miller}}]{Oke1995}
{Oke}, J.~B., {et~al.} 1995, \pasp, 107, 375

\bibitem[{{Osterbrock} \& {Ferland}(2006)}]{Osterbrock:2006}
{Osterbrock}, D.~E., \& {Ferland}, G.~J. 2006, {Astrophysics of gaseous nebulae
  and active galactic nuclei}, ed. {Osterbrock, D.~E.~\& Ferland, G.~J.}

\bibitem[{{Peng} {et~al.}(2011){Peng}, {Chen}, {Gu}, \& {Hu}}]{Peng:2011}
{Peng}, Z.-X., {Chen}, Y.-M., {Gu}, Q.-S., \& {Hu}, C. 2011, Research in
  Astronomy and Astrophysics, 11, 411

\bibitem[{{Piconcelli} {et~al.}(2010){Piconcelli}, {Vignali}, {Bianchi},
  {Mathur}, {Fiore}, {Guainazzi}, {Lanzuisi}, {Maiolino}, \&
  {Nicastro}}]{Piconcelli2010}
{Piconcelli}, E., {et~al.} 2010, ArXiv e-prints

\bibitem[{{Probst} {et~al.}(2008){Probst}, {George}, {Daly}, {Don}, \&
  {Ellis}}]{Probst:2008}
{Probst}, R.~G., {George}, J.~R., {Daly}, P.~N., {Don}, K., \& {Ellis}, M.
  2008, in Presented at the Society of Photo-Optical Instrumentation Engineers
  (SPIE) Conference, Vol. 7014, Society of Photo-Optical Instrumentation
  Engineers (SPIE) Conference Series

\bibitem[{{Probst} {et~al.}(2004){Probst}, {Gaughan}, {Abraham}, {Andrew},
  {Daly}, {Hileman}, {Hunten}, {Liang}, {Merrill}, {Repp}, \&
  {Shaw}}]{Probst:2004}
{Probst}, R.~G., {et~al.} 2004, in Presented at the Society of Photo-Optical
  Instrumentation Engineers (SPIE) Conference, Vol. 5492, Society of
  Photo-Optical Instrumentation Engineers (SPIE) Conference Series, ed.
  {A.~F.~M.~Moorwood \& M.~Iye}, 1716--1724

\bibitem[{{Richstone} {et~al.}(1998){Richstone}, {Ajhar}, {Bender}, {Bower},
  {Dressler}, {Faber}, {Filippenko}, {Gebhardt}, {Green}, {Ho}, {Kormendy},
  {Lauer}, {Magorrian}, \& {Tremaine}}]{Richstone1998}
{Richstone}, D., {et~al.} 1998, \nat, 395, A14+

\bibitem[{{Rodriguez} {et~al.}(2006){Rodriguez}, {Taylor}, {Zavala}, {Peck},
  {Pollack}, \& {Romani}}]{Rodriguez:2006}
{Rodriguez}, C., {Taylor}, G.~B., {Zavala}, R.~T., {Peck}, A.~B., {Pollack},
  L.~K., \& {Romani}, R.~W. 2006, \apj, 646, 49

\bibitem[{{Rosario} {et~al.}(2011){Rosario}, {McGurk}, {Max}, {Shields},
  {Smith}, \& {Ammons}}]{Rosario:2011}
{Rosario}, D.~J., {McGurk}, R.~C., {Max}, C.~E., {Shields}, G.~A., {Smith},
  K.~L., \& {Ammons}, S.~M. 2011, ArXiv e-prints

\bibitem[{{Rosario} {et~al.}(2010){Rosario}, {Shields}, {Taylor}, {Salviander},
  \& {Smith}}]{Rosario:2010}
{Rosario}, D.~J., {Shields}, G.~A., {Taylor}, G.~B., {Salviander}, S., \&
  {Smith}, K.~L. 2010, \apj, 716, 131

\bibitem[{{Schlegel} {et~al.}(1998){Schlegel}, {Finkbeiner}, \&
  {Davis}}]{Schlegel98}
{Schlegel}, D.~J., {Finkbeiner}, D.~P., \& {Davis}, M. 1998, ApJ, 500, 525

\bibitem[{{Schmitt} \& {Kinney}(1996)}]{Schmitt:1996}
{Schmitt}, H.~R., \& {Kinney}, A.~L. 1996, \apj, 463, 498

\bibitem[{{Schneider} {et~al.}(2007){Schneider}, {Hall}, {Richards}, {Strauss},
  {Vanden Berk}, {Anderson}, {Brandt}, {Fan}, {Jester}, {Gray}, {Gunn},
  {SubbaRao}, {Thakar}, {Stoughton}, {Szalay}, {Yanny}, {York}, {Bahcall},
  {Barentine}, {Blanton}, {Brewington}, {Brinkmann}, {Brunner}, {Castander},
  {Csabai}, {Frieman}, {Fukugita}, {Harvanek}, {Hogg}, {Ivezi{\'c}}, {Kent},
  {Kleinman}, {Knapp}, {Kron}, {Krzesi{\'n}ski}, {Long}, {Lupton}, {Nitta},
  {Pier}, {Saxe}, {Shen}, {Snedden}, {Weinberg}, \& {Wu}}]{Schneider07}
{Schneider}, D.~P., {et~al.} 2007, AJ, 134, 102

\bibitem[{{Shen} {et~al.}(2011){Shen}, {Liu}, {Greene}, \&
  {Strauss}}]{Shen:2011}
{Shen}, Y., {Liu}, X., {Greene}, J.~E., \& {Strauss}, M.~A. 2011, \apj, 735, 48

\bibitem[{{Shields} {et~al.}(2009{\natexlab{a}}){Shields}, {Bonning}, \&
  {Salviander}}]{Shields2009}
{Shields}, G.~A., {Bonning}, E.~W., \& {Salviander}, S. 2009{\natexlab{a}},
  \apj, 696, 1367

\bibitem[{{Shields} {et~al.}(2009{\natexlab{b}}){Shields}, {Rosario}, {Smith},
  {Bonning}, {Salviander}, {Kalirai}, {Strickler}, {Ramirez-Ruiz}, {Dutton},
  {Treu}, \& {Marshall}}]{Shields09}
{Shields}, G.~A., {et~al.} 2009{\natexlab{b}}, ApJ, 707, 936

\bibitem[{{Smith} {et~al.}(2010){Smith}, {Shields}, {Bonning}, {McMullen},
  {Rosario}, \& {Salviander}}]{Smith2010}
{Smith}, K.~L., {Shields}, G.~A., {Bonning}, E.~W., {McMullen}, C.~C.,
  {Rosario}, D.~J., \& {Salviander}, S. 2010, \apj, 716, 866

\bibitem[{{Smith} {et~al.}(2011){Smith}, {Shields}, {Salviander}, {Stevens}, \&
  {Rosario}}]{Smith:2011}
{Smith}, K.~L., {Shields}, G.~A., {Salviander}, S., {Stevens}, A.~C., \&
  {Rosario}, D.~J. 2011, ArXiv e-prints

\bibitem[{{Springel} {et~al.}(2005){Springel}, {Di Matteo}, \&
  {Hernquist}}]{Springel:2005}
{Springel}, V., {Di Matteo}, T., \& {Hernquist}, L. 2005, \mnras, 361, 776

\bibitem[{{Stern} {et~al.}(2005){Stern}, {Eisenhardt}, {Gorjian}, {Kochanek},
  {Caldwell}, {Eisenstein}, {Brodwin}, {Brown}, {Cool}, {Dey}, {Green},
  {Jannuzi}, {Murray}, {Pahre}, \& {Willner}}]{Stern2005}
{Stern}, D., {et~al.} 2005, \apj, 631, 163

\bibitem[{{Strateva} {et~al.}(2003){Strateva}, {Strauss}, {Hao}, {Schlegel},
  {Hall}, {Gunn}, {Li}, {Ivezi{\'c}}, {Richards}, {Zakamska}, {Voges},
  {Anderson}, {Lupton}, {Schneider}, {Brinkmann}, \& {Nichol}}]{Strateva03}
{Strateva}, I.~V., {et~al.} 2003, AJ, 126, 1720

\bibitem[{{Tadhunter} {et~al.}(1988){Tadhunter}, {Fosbury}, {di Serego
  Alighieri}, {Bland}, {Danziger}, {Goss}, {McAdam}, \&
  {Snijders}}]{Tadhunter:1988}
{Tadhunter}, C.~N., {Fosbury}, R.~A.~E., {di Serego Alighieri}, S., {Bland},
  J., {Danziger}, I.~J., {Goss}, W.~M., {McAdam}, W.~B., \& {Snijders},
  M.~A.~J. 1988, \mnras, 235, 403

\bibitem[{{Tang} \& {Grindlay}(2009)}]{TG09}
{Tang}, S., \& {Grindlay}, J. 2009, ApJ, 704, 1189

\bibitem[{{van Dam} {et~al.}(2004){van Dam}, {Le Mignant}, \&
  {Macintosh}}]{vanDam2004}
{van Dam}, M.~A., {Le Mignant}, D., \& {Macintosh}, B.~A. 2004, \ao, 43, 5458

\bibitem[{{van Ojik} {et~al.}(1996){van Ojik}, {Roettgering}, {Carilli},
  {Miley}, {Bremer}, \& {Macchetto}}]{vanOjik:1996}
{van Ojik}, R., {Roettgering}, H.~J.~A., {Carilli}, C.~L., {Miley}, G.~K.,
  {Bremer}, M.~N., \& {Macchetto}, F. 1996, \aap, 313, 25

\bibitem[{{Villar-Mart{\'{\i}}n} {et~al.}(2000){Villar-Mart{\'{\i}}n},
  {Alonso-Herrero}, {di Serego Alighieri}, \& {Vernet}}]{Villar-Martin:2000}
{Villar-Mart{\'{\i}}n}, M., {Alonso-Herrero}, A., {di Serego Alighieri}, S., \&
  {Vernet}, J. 2000, \aaps, 147, 291

\bibitem[{{Villar-Mart{\'{\i}}n} {et~al.}(2007){Villar-Mart{\'{\i}}n},
  {S{\'a}nchez}, {Humphrey}, {Dijkstra}, {di Serego Alighieri}, {De Breuck}, \&
  {Gonz{\'a}lez Delgado}}]{Villar-Martin:2007}
{Villar-Mart{\'{\i}}n}, M., {S{\'a}nchez}, S.~F., {Humphrey}, A., {Dijkstra},
  M., {di Serego Alighieri}, S., {De Breuck}, C., \& {Gonz{\'a}lez Delgado}, R.
  2007, \mnras, 378, 416

\bibitem[{{Villar-Mart{\'{\i}}n} {et~al.}(2010){Villar-Mart{\'{\i}}n},
  {Tadhunter}, {P{\'e}rez}, {Humphrey}, {Mart{\'{\i}}nez-Sansigre}, {Delgado},
  \& {P{\'e}rez-Torres}}]{Villar-Martin:2010}
{Villar-Mart{\'{\i}}n}, M., {Tadhunter}, C., {P{\'e}rez}, E., {Humphrey}, A.,
  {Mart{\'{\i}}nez-Sansigre}, A., {Delgado}, R.~G., \& {P{\'e}rez-Torres}, M.
  2010, \mnras, 407, L6

\bibitem[{{Vivek} {et~al.}(2009){Vivek}, {Srianand}, {Noterdaeme}, {Mohan}, \&
  {Kuriakosde}}]{Vivek2009}
{Vivek}, M., {Srianand}, R., {Noterdaeme}, P., {Mohan}, V., \& {Kuriakosde},
  V.~C. 2009, \mnras, 400, L6

\bibitem[{{Wang} {et~al.}(2009){Wang}, {Chen}, {Hu}, {Mao}, {Zhang}, \&
  {Bian}}]{Wang2009}
{Wang}, J., {Chen}, Y., {Hu}, C., {Mao}, W., {Zhang}, S., \& {Bian}, W. 2009,
  \apjl, 705, L76

\bibitem[{{Wang} {et~al.}(2004){Wang}, {Malhotra}, {Rhoads}, {Brown}, {Dey},
  {Heckman}, {Jannuzi}, {Norman}, {Tiede}, \& {Tozzi}}]{Wang04}
{Wang}, J.~X., {et~al.} 2004, AJ, 127, 213

\bibitem[{{Xu} \& {Komossa}(2009)}]{XK09}
{Xu}, D., \& {Komossa}, S. 2009, ApJ, 705, L20

\bibitem[{{Zhou} {et~al.}(2004){Zhou}, {Wang}, {Zhang}, {Dong}, \&
  {Li}}]{Zhou04}
{Zhou}, H., {Wang}, T., {Zhang}, X., {Dong}, X., \& {Li}, C. 2004, ApJ, 604,
  L33

\bibitem[{{Zirm} {et~al.}(2005){Zirm}, {Overzier}, {Miley}, {Blakeslee},
  {Clampin}, {De Breuck}, {Demarco}, {Ford}, {Hartig}, {Homeier},
  {Illingworth}, {Martel}, {R{\"o}ttgering}, {Venemans}, {Ardila}, {Bartko},
  {Ben{\'{\i}}tez}, {Bouwens}, {Bradley}, {Broadhurst}, {Brown}, {Burrows},
  {Cheng}, {Cross}, {Feldman}, {Franx}, {Golimowski}, {Goto}, {Gronwall},
  {Holden}, {Infante}, {Kimble}, {Krist}, {Lesser}, {Mei}, {Menanteau},
  {Meurer}, {Motta}, {Postman}, {Rosati}, {Sirianni}, {Sparks}, {Tran},
  {Tsvetanov}, {White}, \& {Zheng}}]{Zirm:2005}
{Zirm}, A.~W., {et~al.} 2005, \apj, 630, 68

\end{thebibliography}
\end{document}